\documentclass[unsortedaddress,aps,preprint]{revtex4}
\usepackage{graphicx}
\usepackage{dcolumn}
\usepackage{amsmath}
\usepackage{amssymb}
\usepackage{amsfonts}
\usepackage{tikz}
\usepackage{bm}
\usepackage{color}
\usepackage[normalem]{ulem}
\usepackage{hyperref}
\hypersetup{colorlinks=true,linkcolor=blue,urlcolor=magenta}
\newcommand{\be}{\begin{equation}}
\newcommand{\ee}{\end{equation}}
\newcommand{\ba}{\begin{eqnarray}}
\newcommand{\ea}{\end{eqnarray}}
\newcommand*\circled[1]{\tikz[baseline=(char.base)]{\node[shape=circle,draw,inner sep=0.3pt] (char) {#1};}}

\begin{document}
\title{Classical and Quantum Gases on a Semiregular Mesh}
\author{Davide De Gregorio $^1$ and Santi Prestipino $^1$\footnote{Email: {\tt sprestipino@unime.it}}}
\affiliation{$^1$Universit\`a degli Studi di Messina, Dipartimento di Scienze Matematiche ed Informatiche, Scienze Fisiche e Scienze della Terra, Viale F. Stagno d'Alcontres 31, 98166 Messina, Italy}

\begin{abstract}The main objective of a statistical mechanical calculation is drawing the phase diagram of a many-body system. In this respect, discrete systems offer the clear advantage over continuum systems of an easier enumeration of microstates, though at the cost of added abstraction. With this in mind, we examine a system of particles living on the vertices of the (biscribed) pentakis dodecahedron, using different couplings for first and second neighbor particles to induce a competition between icosahedral and dodecahedral orders. After working out the phases of the model at zero temperature, we carry out Metropolis Monte Carlo simulations at finite temperature, highlighting the existence of smooth transitions between distinct ``phases''. The sharpest of these crossovers are characterized by hysteretic behavior near zero temperature, which reveals a bottleneck issue for Metropolis dynamics in state space. Next, we introduce the quantum (Bose-Hubbard) counterpart of the previous model and calculate its phase diagram at zero and finite temperatures using the decoupling approximation. We thus uncover, in addition to Mott insulating ``solids'', also the existence of supersolid ``phases'' which progressively shrink as the system is heated up. We argue that a quantum system of the kind described here can be realized with programmable holographic optical tweezers.
\end{abstract}
\maketitle
\section{Introduction}

Investigating the behavior of a many-particle system has an undeniable charm: despite microscopic interactions are undirected, various forms of self-organization (``order'') can develop at the macroscale. In the last century, countless examples of emergent order have been described, each with its own practical realization, and many more can be devised by exploring through theory physical situations that are somehow atypical. These indications can stimulate new experimental work or simply be aimed to clarify and expand the scope of the theory itself.

A way to produce novel, unconventional phase behaviors is to consider many-body systems under geometric constraints, since local interactions are frustrated and unusual ground states then appear. A classic example is a (finite) system of hard particles confined in the surface of a sphere~\cite{Post}. The sphere topology forces an excess of fivefold coordinated particles over sevenfold ones, leading to high-density packings with defects~\cite{Prestipino,Prestipino2,Prestipino3,Vest,Guerra,Franzini,Dlamini}. We note that bosonic atoms confined in thin spherical shells~\cite{Prestipino4} have already been realized~\cite{Zobay,Garraway} and are currently studied in microgravity~\cite{Elliott,Lundblad}. In other cases, frustration is directly embodied in the interaction law --- like in spin glasses or in antiferromagnets on a triangular lattice~\cite{Wannier,Toulouse}.

In this paper, we consider a discrete system of particles (``lattice gas'') on a spherical mesh of points, which is chosen such that a rich interplay arises between distinct ``phases'' having the symmetries of a Platonic solid. Clearly, on a finite mesh well-definite phases only exist at zero temperature ($T=0$), since for $T>0$ any phase transition will be smeared out, i.e., replaced by a smooth crossover region. Using a finite mesh, we greatly reduce the computational effort without however making the phase behavior trivial.

If a toy model of classical particles on a finite mesh may look somewhat artificial and hardly corresponds to a real-world system, its quantum counterpart might be different. The last decades have witnessed a considerable progress in the manipulation of quantum atoms at low temperature, opening the way to a systematic study of correlation effects in many-body systems~\cite{Bloch,Amico,Jaksch,Greiner}. While optical lattices~\cite{Windpassinger} are routinely employed in numerous laboratories worldwide as a tool for confinement of quantum atoms, in the last few years a laser technology has been invented, based on the use of optical tweezers~\cite{Barredo,Browaeys}, which allows virtually any type of structure (not necessarily a lattice) to be realized with cold atoms. We are thus encouraged to consider the quantum (Bose-Hubbard) counterpart of the lattice gas on a spherical mesh, with the explicit purpose to compare their thermal behaviors. In particular, we devise a quantum variational theory that predicts supersolid phases and returns the results of a classical mean-field theory when quantum tunneling is precluded.

The rest of the paper is organized as follows. In Sec.\,2 we introduce our model and the methods used to investigate its phase behavior. Next, we present our results, first at zero temperature (Sec.\,2.1) and then at finite temperatures (Sec.\,2.2). In Sec. 3 we deal with the quantum extension of the model in Sec.\,2.  Using the decoupling approximation, we not only work out the ground-state diagram (Sec.\,3.1) but also a few finite-temperature properties (Sec.\,3.2). Lastly, we give our concluding remarks in Sec.\,4.

\section{Lattice-gas models on a spherical mesh}

As anticipated in the Introduction, we hereafter explore the possibility of unusual orderings in a system of particles occupying the nodes of a spherical mesh, chosen to be sufficiently regular that some polyhedral (Platonic) arrangement can occur.

%
%
\begin{figure}
\centering
\includegraphics[width=8cm]{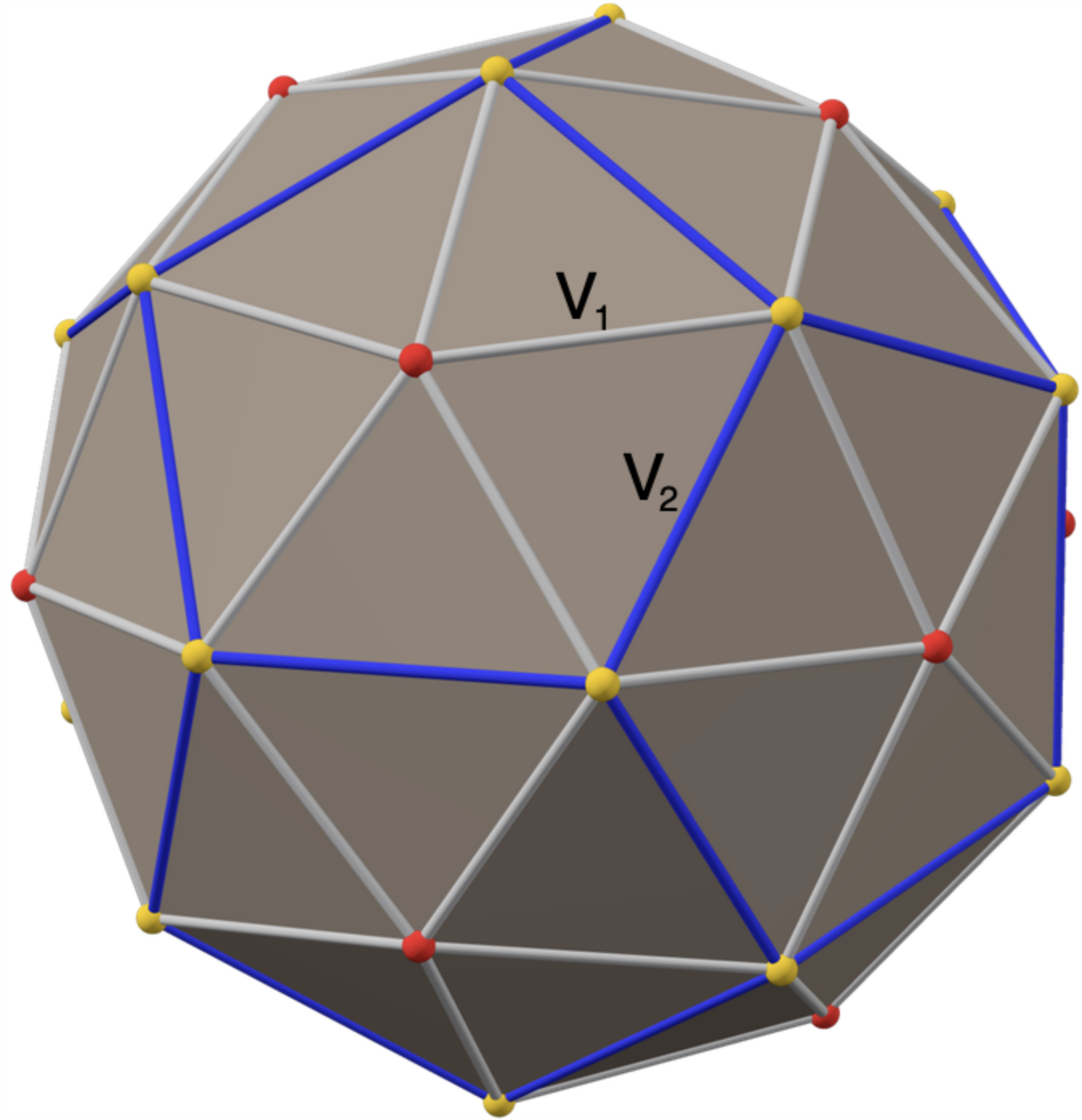}
\caption{The pentakis dodecahedron has 12 icosahedral vertices (red dots) and 20 dodecahedral vertices (yellow dots). There are five distinct ways to choose eight yellow dots forming a cube --- then, the other 12 dodecahedral vertices are said to form a ``co-cube''. The short edges of the PD mesh are colored in grey and the long edges in blue. The couplings entering the model Hamiltonian (\ref{eq-1}) are indicated.}
\end{figure}

We focus our attention on the pentakis dodecahedron~\cite{visualpolyhedra} (PD, see Fig.\,1), a Catalan solid with 32 vertices obtained by augmenting the dodecahedron with 12 right pyramids on its pentagonal faces, in such a way that the resulting polyhedron is dual to the truncated icosahedron (clearly, the pyramidal apices form the vertices of an icosahedron). Even though the PD is not inscribable, implying that no spherical mesh can be drawn from its vertices, it is straightforward to obtain a biscribed solid by a small distortion of the PD that preserves its connectivity properties and its full icosahedral symmetry. A biscribed solid is any convex polyhedron that has concentric circumscribed and inscribed spheres, where the sphere center is also the centroid of the vertices. The five Platonic solids are biscribed solids, but none of the Archimedean or Catalan solids are. As for the PD, it suffices to adjust the height of the pentagonal pyramids only slightly to force all the vertices to be on the same sphere. The outcome of this construction is the biscribed form of the PD. It is this variant of the PD that is considered hereafter.

The PD has 60 faces (isosceles triangles) and 90 edges (60 short and 30 long). We call PD mesh the skeleton of the PD, i.e., the mesh formed by its edges. The PD mesh is the finite analog of a lattice; the nodes of the mesh (i.e., the PD vertices) are its ``sites''. While five edges depart from an icosahedral vertex/site, the number of edges departing from a dodecahedral vertex/site is six (in this sense, icosahedral and dodecahedral sites are ``inequivalent''). Setting the circumscribed radius equal to 1, the short-edge length (i.e., the shortest distance in the mesh) is $\ell_1=\sqrt{30\big(15-\sqrt{15(5+2\sqrt{5})}\big)}/15\simeq 0.64085\ldots$, whereas the long-edge length (the second shortest distance in the mesh) is $\ell_2=\big(\sqrt{15}-\sqrt{3}\big)/3\simeq 0.71364\ldots$

The phase behavior of a lattice-gas model on the PD mesh is better studied in the grand-canonical ensemble. Upon increasing the chemical potential $\mu$ at fixed $T$, the mesh becomes increasingly populated, with the possibility of ``transitions'' between qualitatively distinct arrangements. Denoting $c_i=0,1$ the occupation number of site $i$, we call first (second) neighbors any two sites/particles that are linked by a short (long) edge. According to our nomenclature, an icosahedral site has five first-neighbor sites and no second-neighbor site, whereas a dodecahedral site has three first neighbors and three second neighbors (see Fig.\,1). With these specifications, the grand Hamiltonian of our model reads:
\be
H=V_1\sum_{\langle i,j\rangle}c_ic_j+V_2\sum_{\langle\langle k,l\rangle\rangle}c_kc_l-\mu\sum_ic_i\,,
\label{eq-1}
\ee
where $V_1>0$ ($V_2$) is the coupling between two first (second) neighbor particles. This model can mimic a system of particles adsorbed on a ``substrate'' sculpted like a PD mesh (think of, e.g., the interstices between atoms in a C$_{60}$ molecule), and interacting via a spherically-symmetric potential with a hard core followed, at larger distances, by a soft short-range repulsion. By suitably tuning the ratio $\gamma=V_2/V_1$ between the couplings, we anticipate the existence of a competition between icosahedral order and dodecahedral order at low temperature, with the further possibility of arrangements with intermediate order.

In the following, we use $V_1$ as the unit of energy; in turn, this defines a reduced temperature, $T^*=k_BT/V_1$ ($k_B$ being the Boltzmann constant), and a reduced chemical potential, $\mu^*=\mu/V_1$.

\subsection{Zero-temperature phases}

At zero temperature, the stable phase at fixed $\mu$ is the one minimizing the grand potential $\Omega$. We expect the absolute minimum $\Omega$ to be reached in one of a few microstates/configurations, chosen among those exhibiting a homogeneous occupancy of equivalent sites. To be clear, there are five ways to select --- out of 20 dodecahedral sites --- eight sites forming a cube~\cite{Prestipino5} (similarly, each cube is the union of two tetrahedra). Then, the PD vertices are naturally grouped in three sets of equivalent nodes: we call A the set of icosahedral sites, B any set of cubic sites, and C the set comprising the remaining dodecahedral sites (``co-cubic'' sites). For example, in the icosahedral phase (ICO) only A sites are occupied at $T=0$; in the dodecahedral phase (DOD) only B and C sites are occupied. In our setting, ICO and DOD play a role analogous to two distinct crystalline phases. By a straightforward count of neighbors, the grand potential of the most relevant phases is readily calculated: in an obvious notation, $\Omega_{\rm empty}=0$, $\Omega_{\rm ICO}=-12\mu$, $\Omega_{\rm COC}=6V_2-12\mu$, $\Omega_{\rm DOD}=30V_2-20\mu$, and $\Omega_{\rm full}=60V_1+30V_2-32\mu$. The tenfold-degenerate tetrahedral phase (TET, $\Omega=-4\mu$) and the fivefold-degenerate cubic phase (CUB, $\Omega=-8\mu$) are never stable, since less stable than ``empty'' or ICO. Further possibilities are configurations where, in addition to icosahedral sites, also a selection of dodecahedral sites are occupied: $\Omega_{\rm ICO+TET}=12V_1-16\mu$, $\Omega_{\rm ICO+CUB}=24V_1-20\mu$, and $\Omega_{\rm ICO+COC}=36V_1+6V_2-24\mu$.

%
%
\begin{figure}
\begin{center}
\includegraphics[angle=-90,width=13cm]{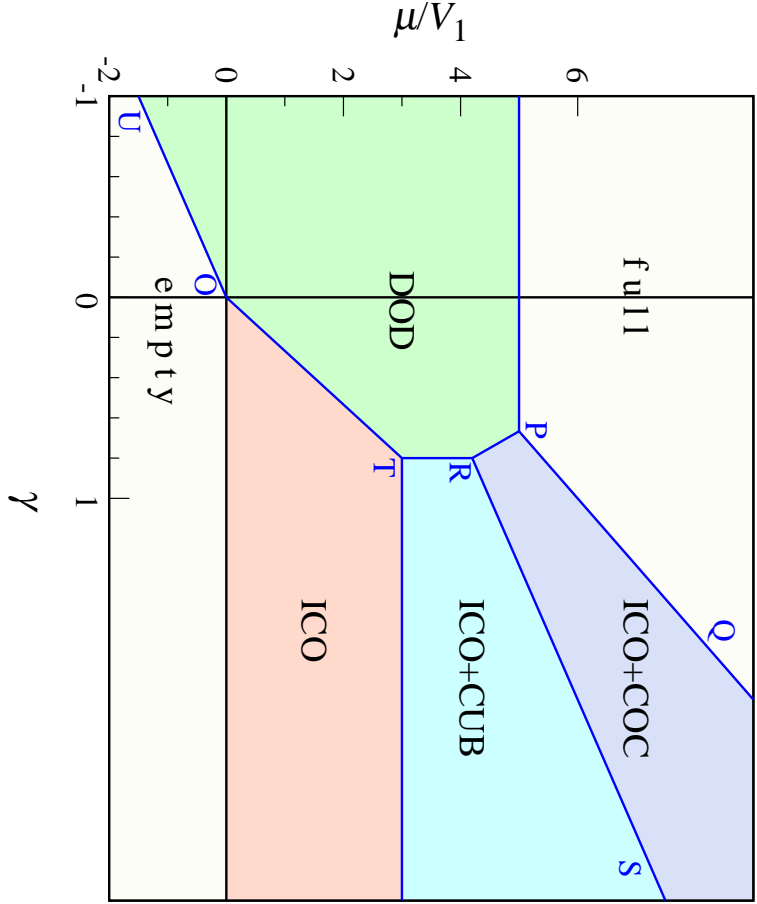}
\end{center}
\caption{Zero-temperature phase diagram of the lattice gas on a PD mesh. The phase boundaries are reported in the main text.}
\label{fig2}
\end{figure}

By a lengthy (though elementary) calculation, we arrive at the $T=0$ phase diagram depicted in Fig.\,2. Here, PQ, PR, and RS are straight lines with equations $\mu^*=3+3\gamma$, $\mu^*=9-6\gamma$, and $\mu^*=3+(3/2)\gamma$, respectively; the two half-lines departing from the origin, namely OT and OU, have equations $\mu^*=(15/4)\gamma$ and $\mu^*=(3/2)\gamma$, respectively. All the phase boundaries are first-order (since the number of particles changes discontinuously from one phase to the other), except for RT (since both the particle number $N$ and the energy $E$ change continuously through it). Looking at Fig.\,2, we note that: i) the ICO phase is only stable for $\gamma>0$, in a region delimited by the $\mu^*=0$ and $\mu^*=3$ lines; ii) the DOD phase is stable in a wide region of the $\gamma$-$\mu$ plane, bounded from the right by $\gamma=4/5$ and from the above by $\mu^*=5$; iii) ICO+CUB and ICO+COC are each stable in an unbounded set of positive $\gamma$ values, and coexist along the RS line; iv) for each $\gamma$, ``empty'' and ``full'' are respectively stable for all values of $\mu$ that are sufficiently small or sufficiently large.

\subsection{Finite-temperature behavior}

For $T>0$ the study of model (\ref{eq-1}) cannot be fully analytical since the number of microstates, $2^{32}$, is huge. Thus, we have resorted to grand-canonical Monte Carlo (MC) simulation for the computation of thermal averages. We employ the standard Metropolis algorithm with single-site moves: at each step of the Markov chain, a move is attempted by changing the state of a randomly selected site $j$ from $c_j$ to $1-c_j$. Calling $\Delta H$ the virtual change in $H$, the move is accepted according to a probability given by $\min\{1,{\rm exp}(-\beta\Delta H)\}$ with $\beta=(k_BT)^{-1}$, as usual. Once equilibrium has been reached, statistical averages are computed over no less than five million MC cycles (one cycle consisting of 32 trial moves).

We monitor a number of equilibrium properties as a function of $\mu$: the number of particles ($N=\langle\sum_ic_i\rangle$) and the energy ($E=\langle H\rangle+\mu N$), together with their self- and cross-correlations; the reduced isothermal compressibility,
\be
\rho k_{\rm B}TK_T=\frac{\langle\delta N^2\rangle}{N}\,\,\,{\rm with}\,\,\,\rho=N/32\,\,{\rm and}\,\,\delta N=\sum_ic_i-N\,;
\label{eq-2}
\ee
and two specific heats, namely
\be
C_\mu=\frac{T}{N}\left.\frac{\partial S}{\partial T}\right|_{V,\mu}\,\,\,{\rm and}\,\,\,
C_N=\frac{T}{N}\left.\frac{\partial S}{\partial T}\right|_{V,N}
\label{eq-3}
\ee
($S$ being the entropy), expressed in terms of grand-canonical averages through Eqs.\,(\ref{a-6}) and (\ref{a-15}) of Appendix A. In addition, we also determine a number of order parameters (OPs, see Appendix B for a definition), in order to establish the nature of the system ``phase'' and the crossover behavior at its boundaries.

For the sake of illustration, take $\gamma=1/2$. For $T=0$ the succession of phases is
\be
{\rm empty}\,\,\stackrel{0}{\longrightarrow}\,\,{\rm ICO}\,\,\stackrel{1.875}{\longrightarrow}\,\,{\rm DOD}\,\,\stackrel{5}{\longrightarrow}\,\,{\rm full}\,.
\label{eq-4}
\ee
We explore the phase behavior of this model at five temperatures, $T^*=0.1,0.2,\ldots 0.5$, and in the $\mu^*$ range from $-2$ to 7. To account for the possibility of hysteresis, we carry out our simulation runs in sequence: for a given value of $\mu^*$, the run is started from the last configuration generated in the previous run at a slightly larger or smaller $\mu^*$. Our results are collected in Figs.\,3 and 4. In the former figure, we plot $N,E,O_{\rm ICO}$, and $O_{\rm DOD}$ as a function of $\mu^*$. As $T$ progressively grows, icosahedral and dodecahedral orders become increasingly weakened, and the crossover region between distinct ``phases'' gets wider and wider. An interesting behavior occurs for $T^*=0.1$, where we observe a large hysteresis loop near the ICO-DOD ``transition''. In other words, despite the absence of sharp transitions for $T>0$, at sufficiently low temperature order can be so robust that we observe hysteresis --- the most characteristic feature of a first-order transition. This is the clue to insufficient sampling of the equilibrium distribution, which can be cured by either substantially increasing the length of the MC trajectory or changing the MC algorithm (see more below).

Curiously enough, hysteresis is only found for one of the three transitions present at $T=0$. To clarify this point, it is instructive to compute the acceptance probability of the Metropolis move(s) that initially drive the system out of one phase into another. The crucial quantity to look at is the ensuing variation in $H$, i.e., $\Delta H=\Delta E-\mu\Delta N$, with $\mu$ given by the transition-point value.

%
%
\begin{figure}
\begin{center}
\includegraphics[width=6.6cm]{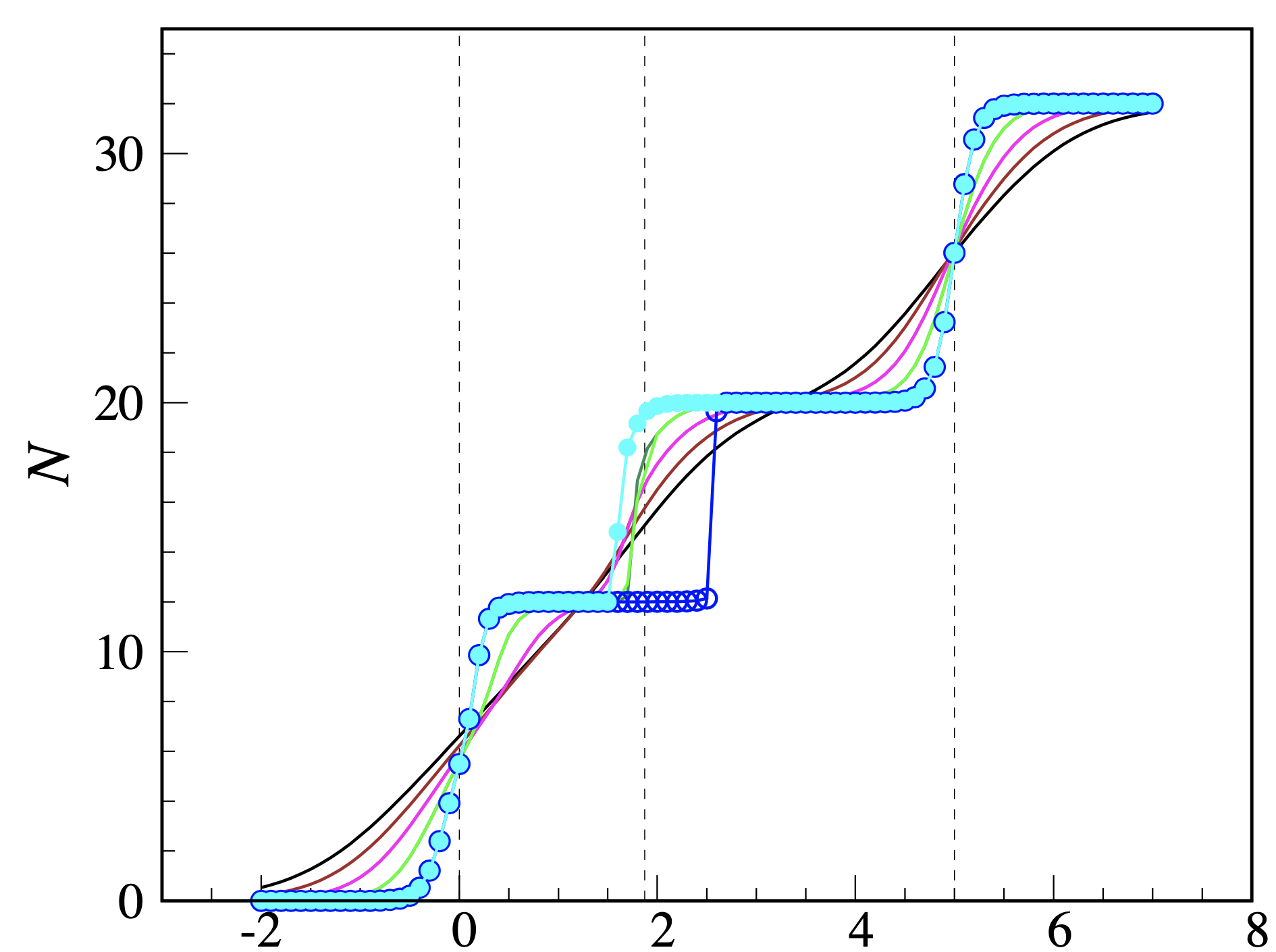}
\includegraphics[width=6.6cm]{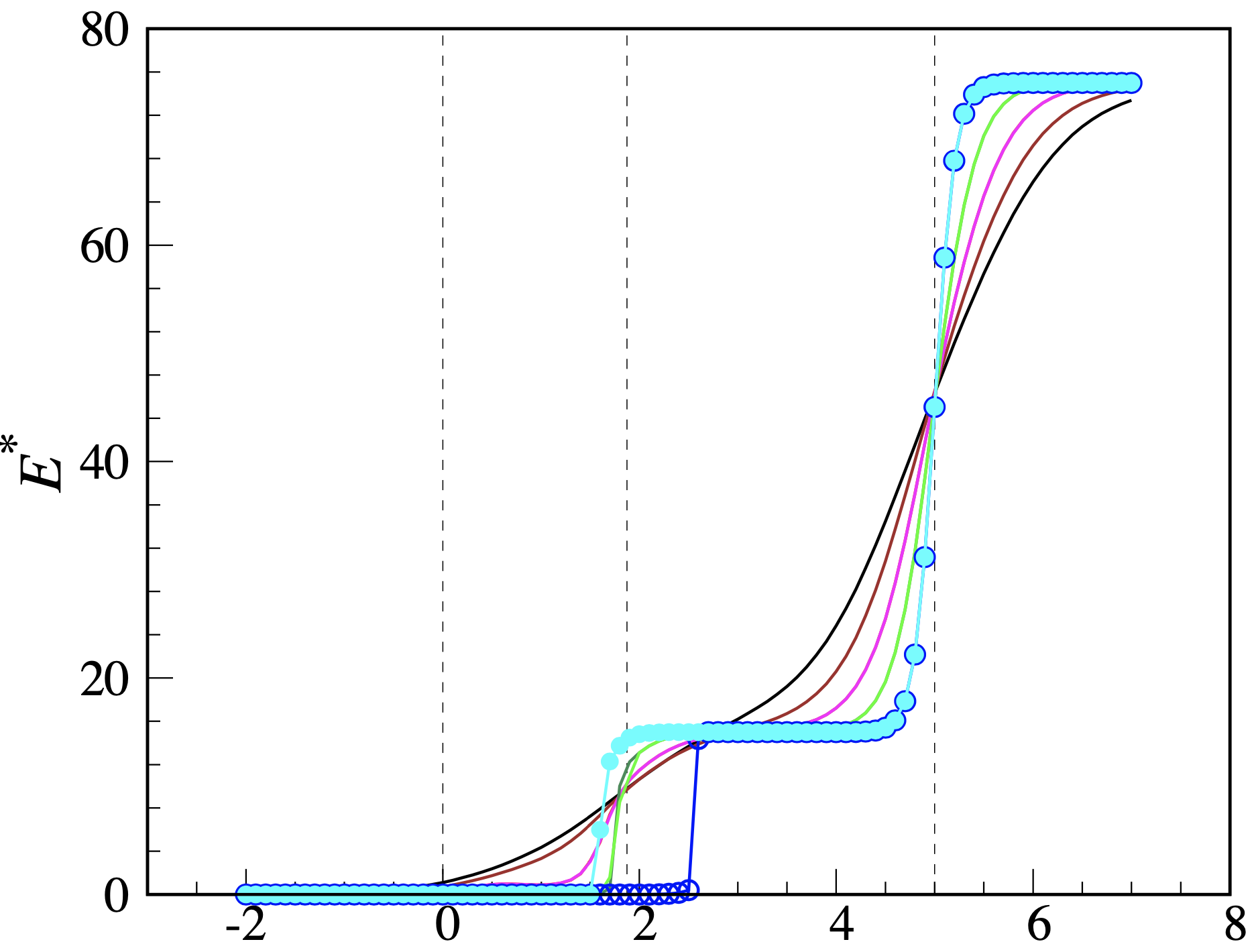} \\
\includegraphics[width=6.6cm]{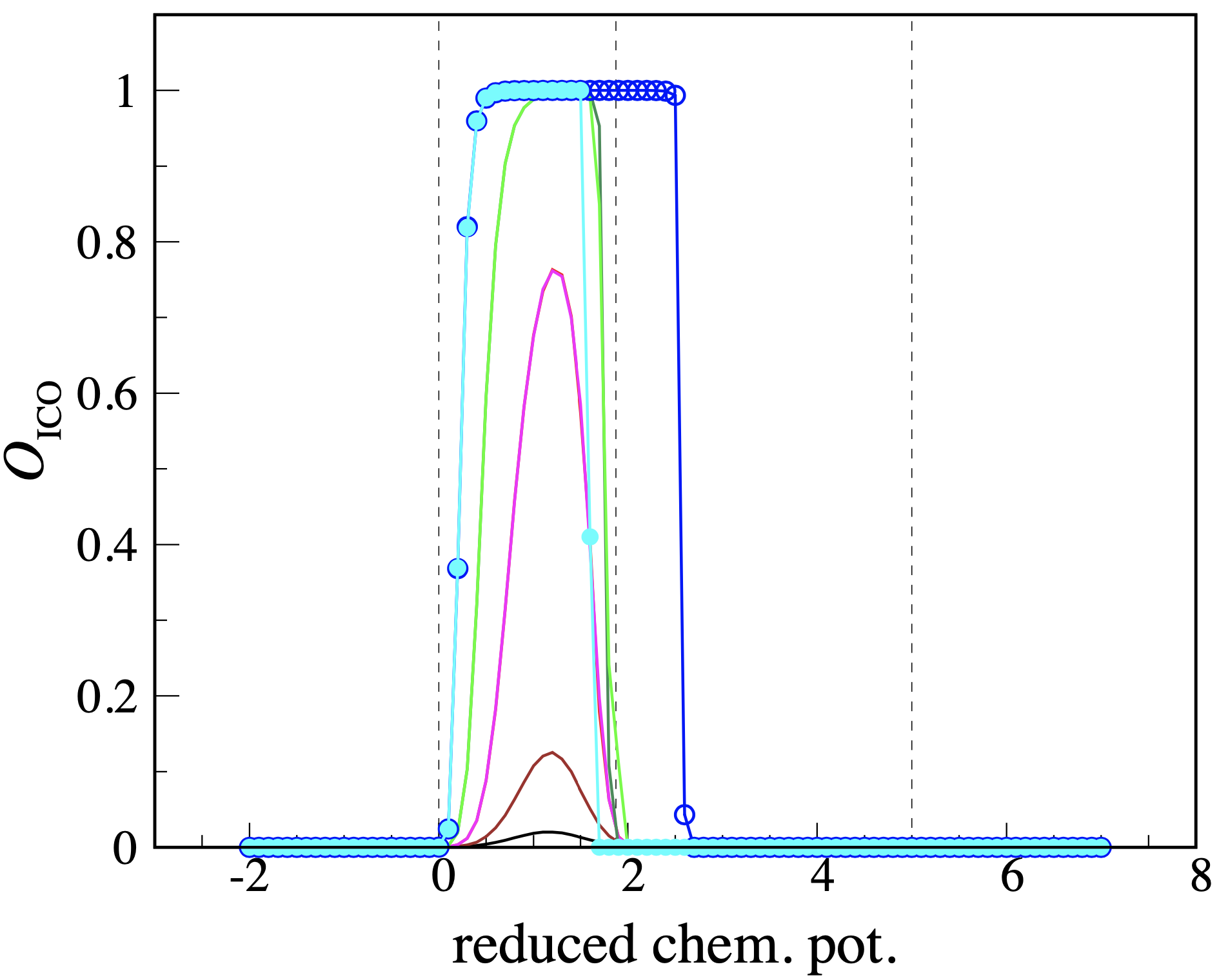}
\includegraphics[width=6.6cm]{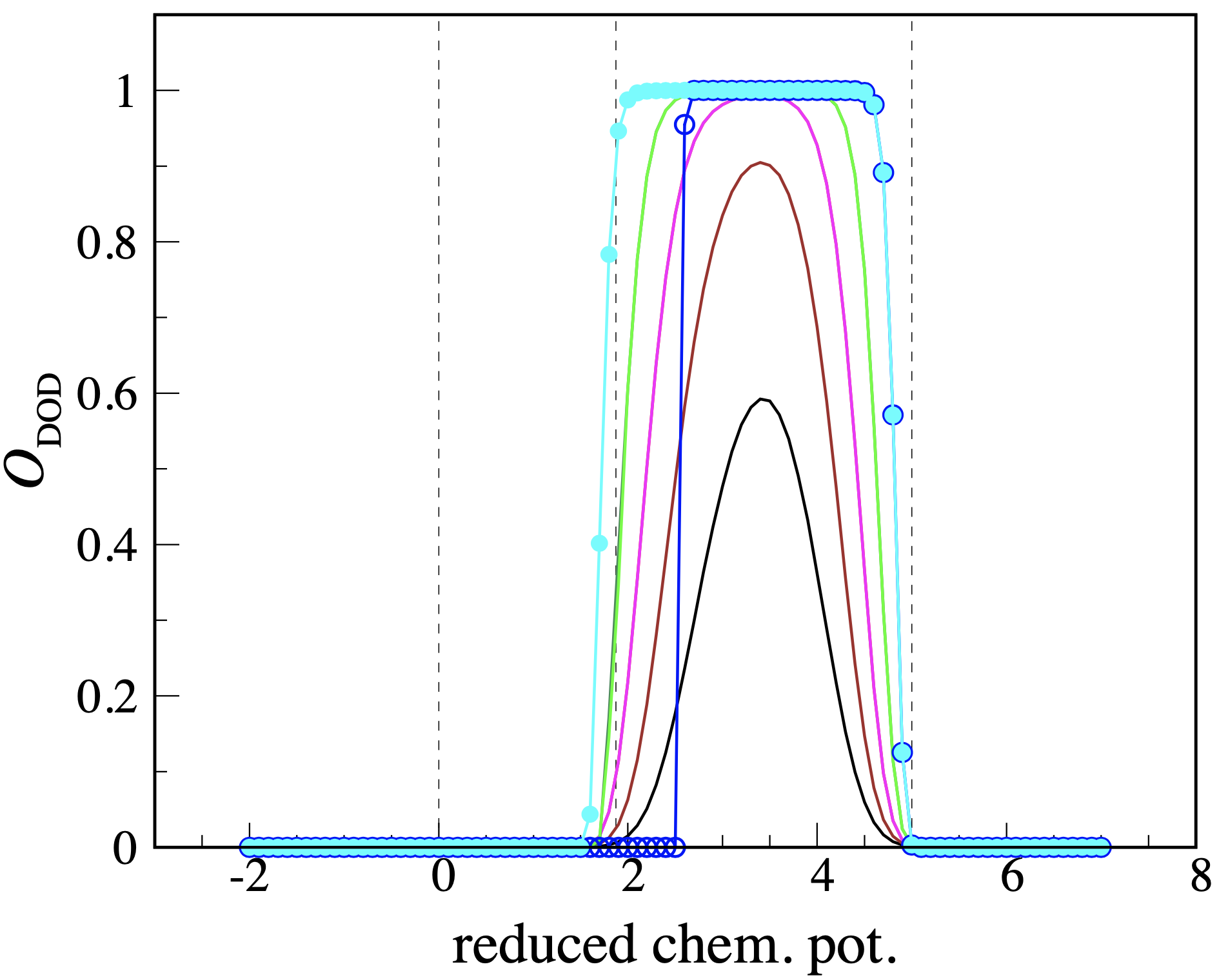}
\end{center}
\caption{Lattice gas on a PD mesh, for $\gamma=1/2$ and $T^*=0.1,0.2,\ldots,0.5$. Top left: number of particles. Top right: reduced energy. Bottom left: icosahedral OP. Bottom right: dodecahedral OP. For the lowest temperatures, MC data refer to two distinct sequences of runs where $\mu^*$ is respectively increased or decreased in steps of 0.1 ($T^*=0.1$, blue and cyan dots; $T^*=0.2$, emerald and green; $T^*=0.3$, red and pink; $T^*=0.4$, brown; and $T^*=0.5$, black). For all temperatures but the lowest one, MC data are reported as lines. Hysteresis is evident for $T^*=0.1$ and barely visible already for $T^*=0.2$.}
\label{fig3}
\end{figure}

Starting from the empty mesh at $\mu=0$, the repeated addition of particles in icosahedral sites occurs with probability 1, since at each step $\Delta E=0$ and $\Delta N=1$. If we start from ICO, the repeated removal of particles again occurs with probability 1, since at each step $\Delta E=0$ and $\Delta N=-1$. This means that for $\mu=0$ the system has equal probability to be in either of the two phases. Hence, no hysteresis would be observed at the empty-ICO transition, as indeed found.

%
%
\begin{figure}
\begin{center}
\includegraphics[width=9cm]{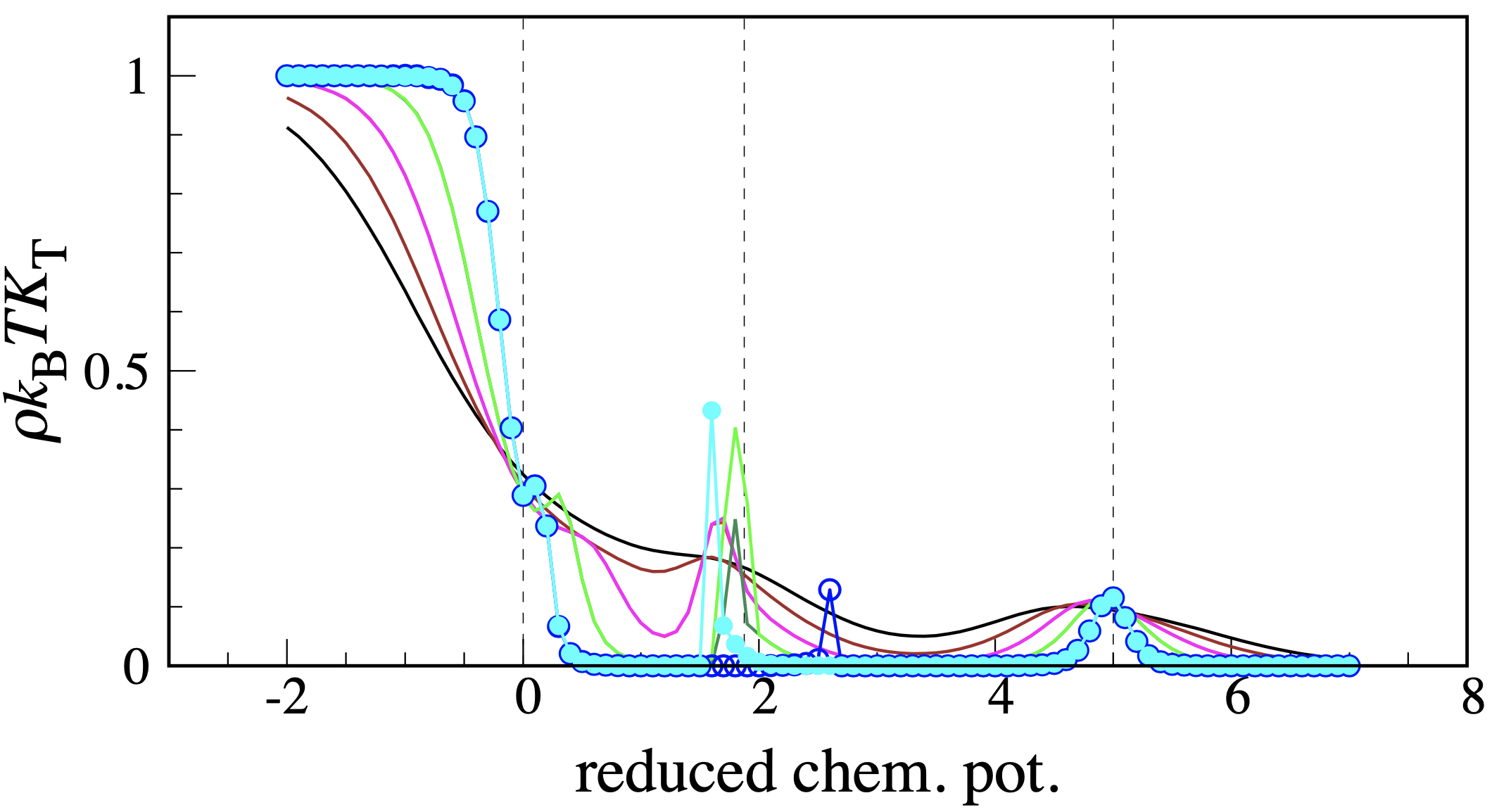} \\
\includegraphics[width=6.4cm]{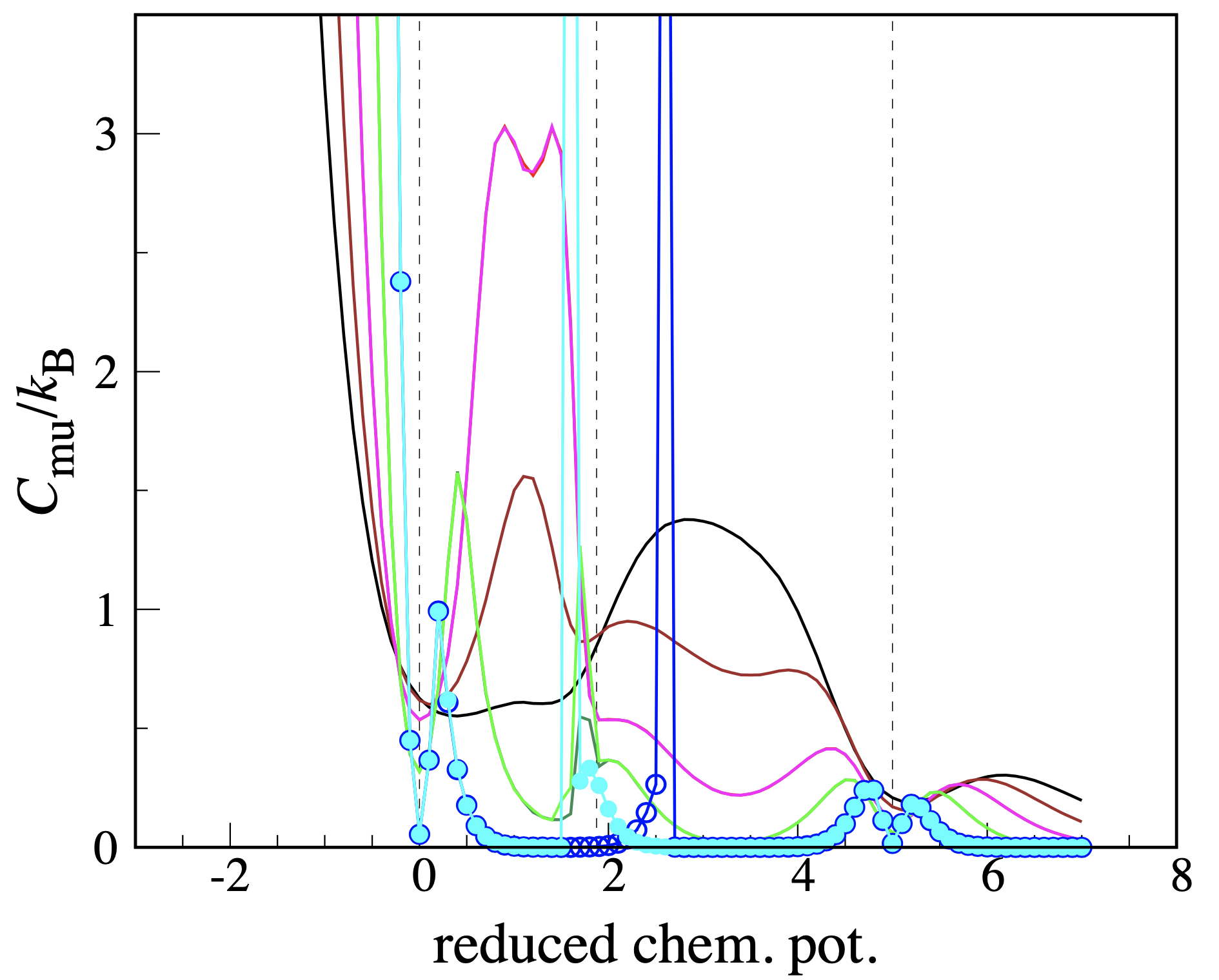}
\includegraphics[width=6.4cm]{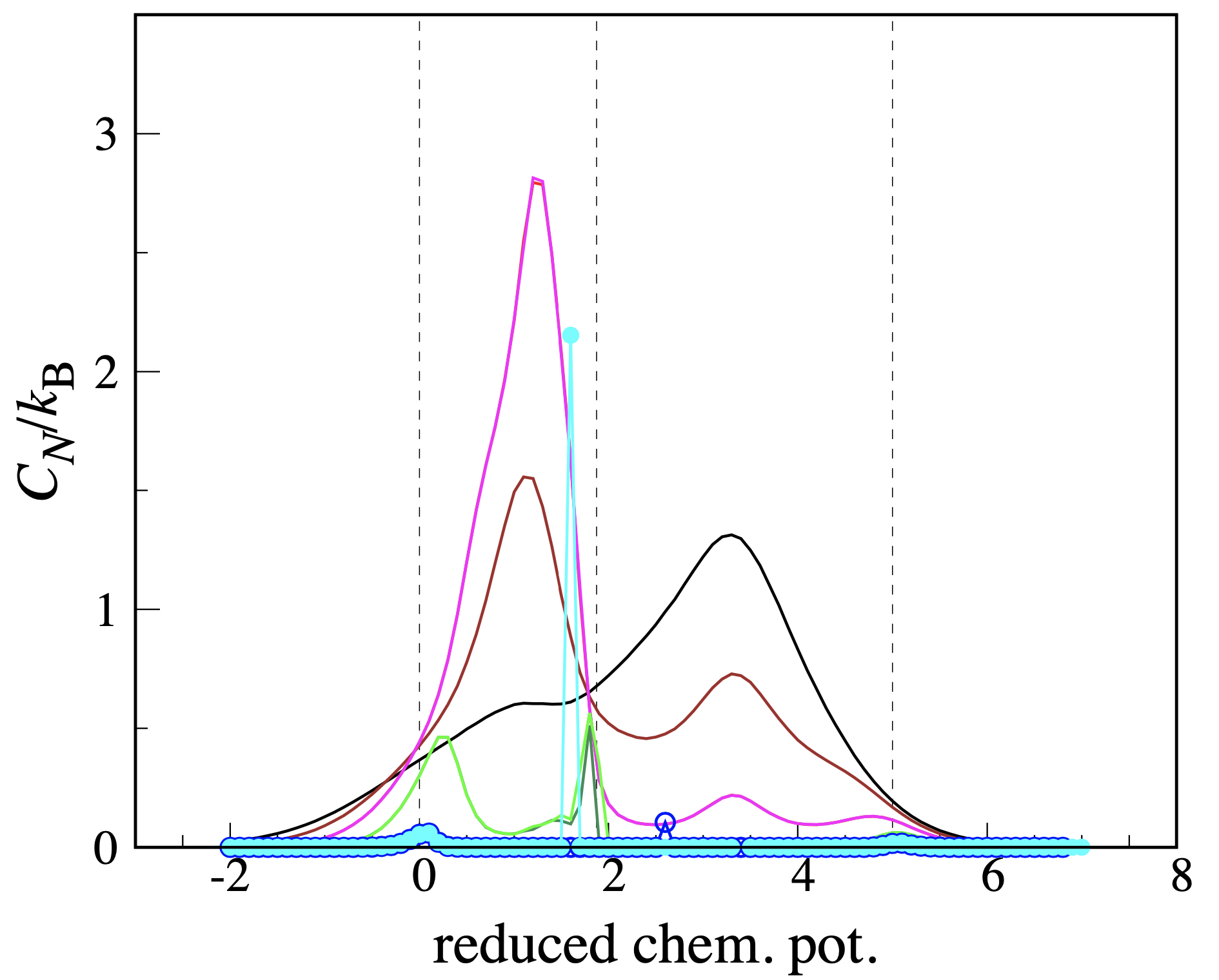}
\end{center}
\caption{Lattice gas on a PD mesh, for $\gamma=1/2$ and $T^*=0.1,0.2,\ldots,0.5$. Top: reduced compressibility. Bottom left: Constant-$\mu$ specific heat. Bottom right: Constant-$N$ specific heat. Symbols and notation as in Fig.\,3.}
\label{fig4}
\end{figure}

Now consider the ICO-DOD transition at $\mu^*=15/8$. Starting from ICO, the cost to annihilate a particle is $\Delta H^*=15/8$ (since $\Delta E=0$); the next step of creating a particle in a dodecahedral site has a minimum cost of $\Delta H^*=31/8$ (since $\Delta E^*\ge 2$). Thus, for $\mu^*=15/8$ there is a free-energy barrier for the transition to DOD, and the system then remains for long in ICO notwithstanding DOD is more stable. If we instead start from DOD, we should first annihilate a particle ($\Delta H^*=3/8$) and then create another particle in an icosahedral site ($\Delta H^*\ge 17/8$). Again, the transition from DOD to ICO is discouraged at low $T$ (though apparently less so than the opposite transition), meaning that for $\mu^*=15/8$ the system is more probably found in DOD than in ICO. As a result, hysteresis will be observed at the ICO-DOD transition.

Lastly, we consider the DOD-full transition at $\mu^*=5$. If we start from DOD, the first step towards ``full'' is creating a particle in an icosahedral site ($\Delta E^*=5$) and the probability for this move is one. If we instead start from ``full'', the cost for annihilating a particle in an icosahedral site is zero again, since $\Delta E^*=-5$. Indeed, no hysteresis is detected at the DOD-full transition.

In Fig.\,4 we plot a few response functions for $\gamma=1/2$. At the lowest temperatures all these quantities exhibit a distinct peak near each $T=0$ transition point, which is where the energy and particle number are subject to the sharpest variations. In the ``empty'' phase the reduced isothermal compressibility is close to 1 (the ideal-gas value); upon heating, every asperity in its profile becomes smoothened until $\rho K_T$ becomes a monotonously decreasing function of $\mu$. As $T$ grows, both specific heats develop a broad maximum inside the ICO and DOD regions. Admittedly, these are the locations where, in the moderately hot system, the fluctuations of energy and particle number are stronger. Eventually, both maxima gradually deflate, the DOD bump being the last to disappear. Finally notice that inside the ``empty'' region the two specific heats have different behaviors at low temperature: while the covariances involving energy are practically zero, the mean square fluctuation of $N$ is of order $N$ (see Eq.\,(\ref{eq-2})). Looking at Eqs.\,(\ref{a-6}) and (\ref{a-15}), we thus conclude that $C_\mu/k_{\rm B}\approx(\beta\mu)^2\gg 1$ and $C_N\approx 0$.

As a second example, consider $\gamma=1$ and $T^*=0.05$ close to $\mu^*=9/2$ --- which is where, at $T=0$, ICO+CUB transforms into ICO+COC. By the same argument put forward before we would conclude that this transition is accompanied by hysteresis at low temperature. However, when plotting $N$ as a function of $\mu$, in a small neighborhood of $\mu^*=9/2$ we see a narrow plateau in the middle between the $N$ levels in the two phases (top-left panel of Fig.\,5), and a similar occurrence is found for $E$ (not shown). This unexpected outcome would suggest the existence of a yet undetected phase near $\mu^*=9/2$, characterized by $N=22$ and $E^*=33$. In fact, when the length of the MC trajectory is increased by a factor of 10 the plateau at $N^*=22$ changes to a more or less smooth crossover between the phases (top-right panel of Fig.\,5). To clarify things better, we have enumerated the microstates with all icosahedral sites occupied, thus confirming that ICO+CUB (ICO+COC) is the stable phase for $\mu^*<9/2$ ($\mu^*>9/2$); instead, exactly for $\mu^*=9/2$ we have counted as many as 240 distinct microstates, all with $N=22$ and $E^*=33$, having the same grand potential as ICO+CUB and ICO+COC. It is the existence of such configurations that makes the transition between ICO+CUB and ICO+COC smoother than expected. We care to stress that this occurrence is rather exceptional; usually, phases are well separated in free energy and compete with each other only in pairs.

%
%
\begin{figure}
\begin{center}
\includegraphics[width=6.4cm]{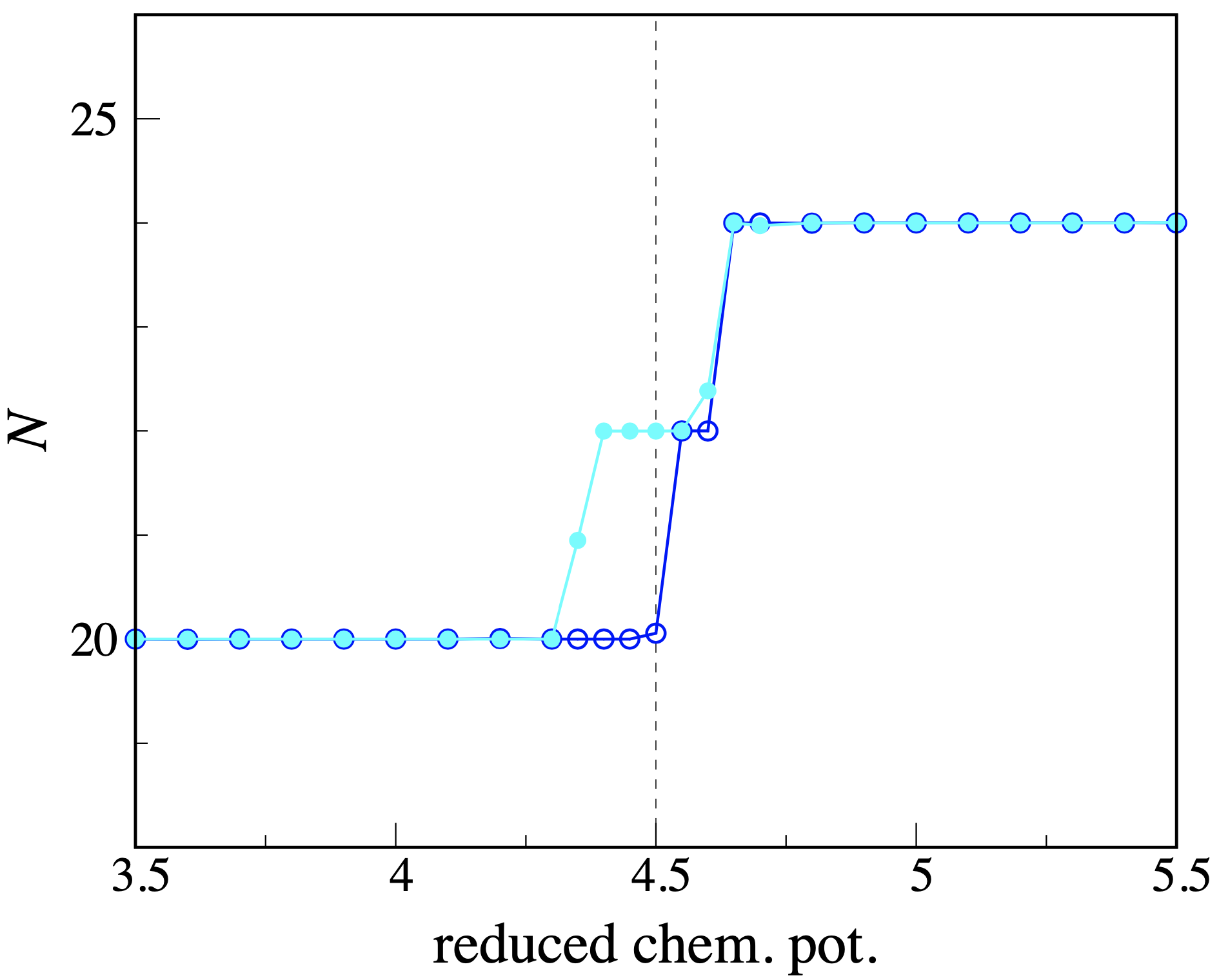}
\includegraphics[width=6.4cm]{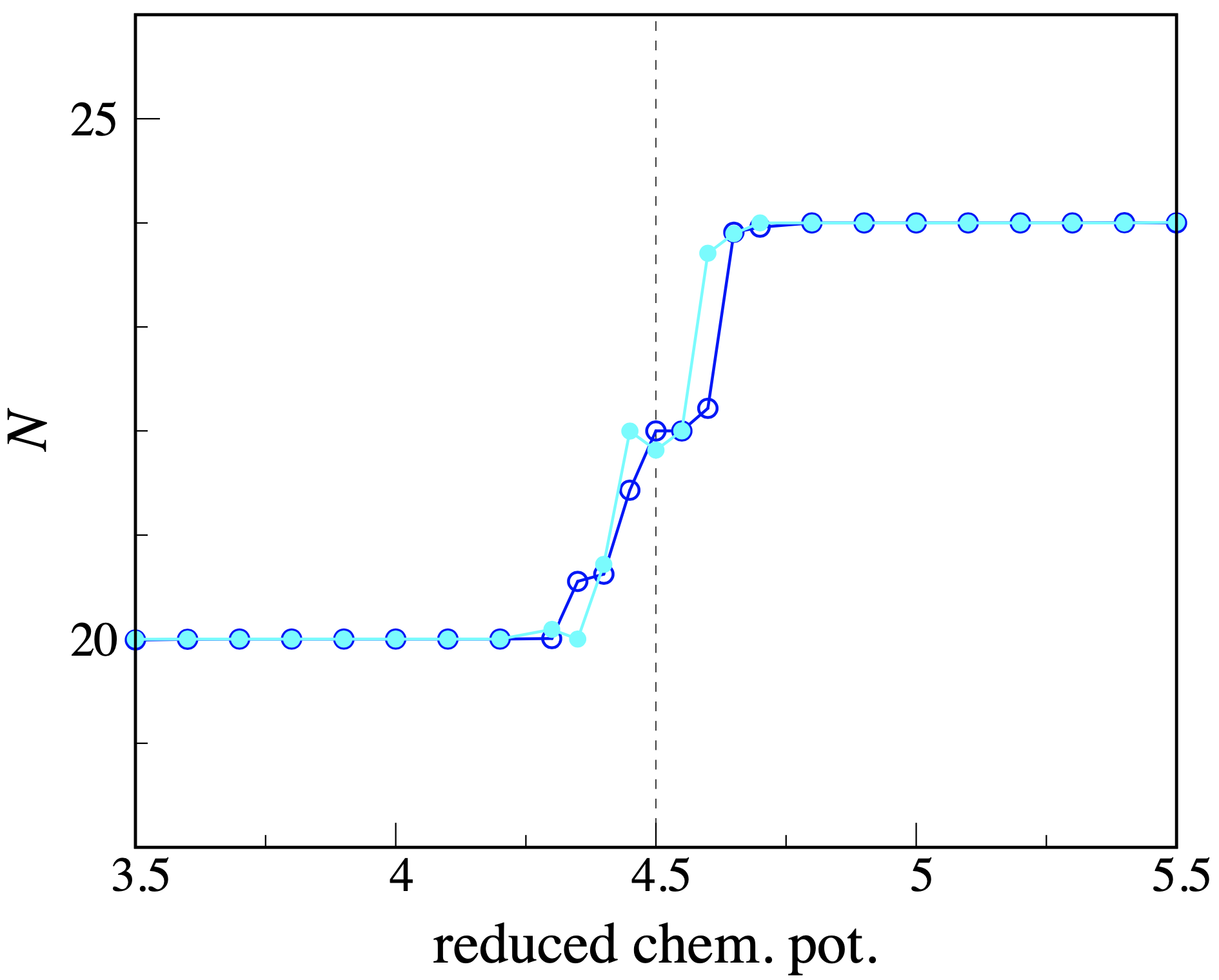} \\
\includegraphics[width=6.4cm]{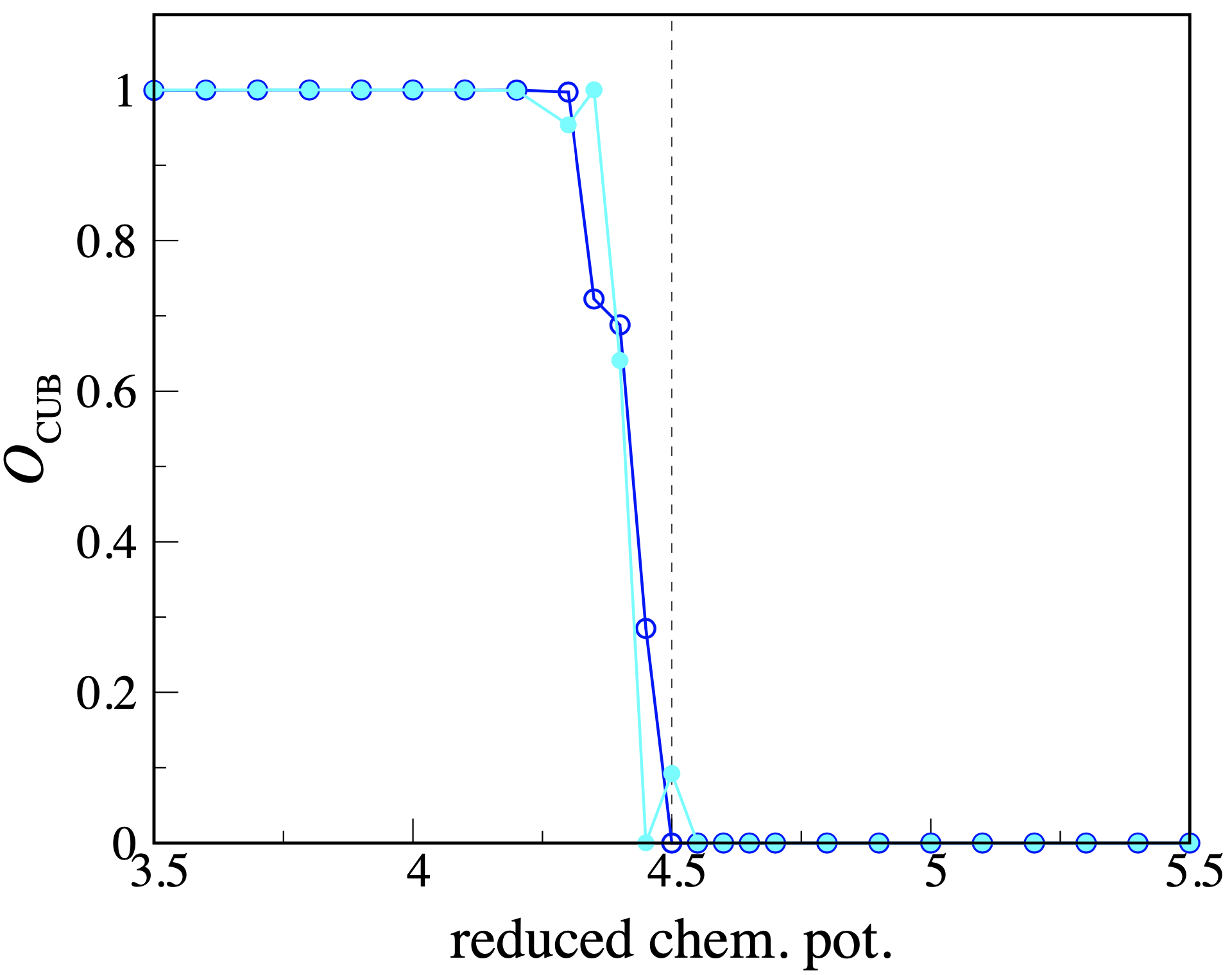}
\includegraphics[width=6.4cm]{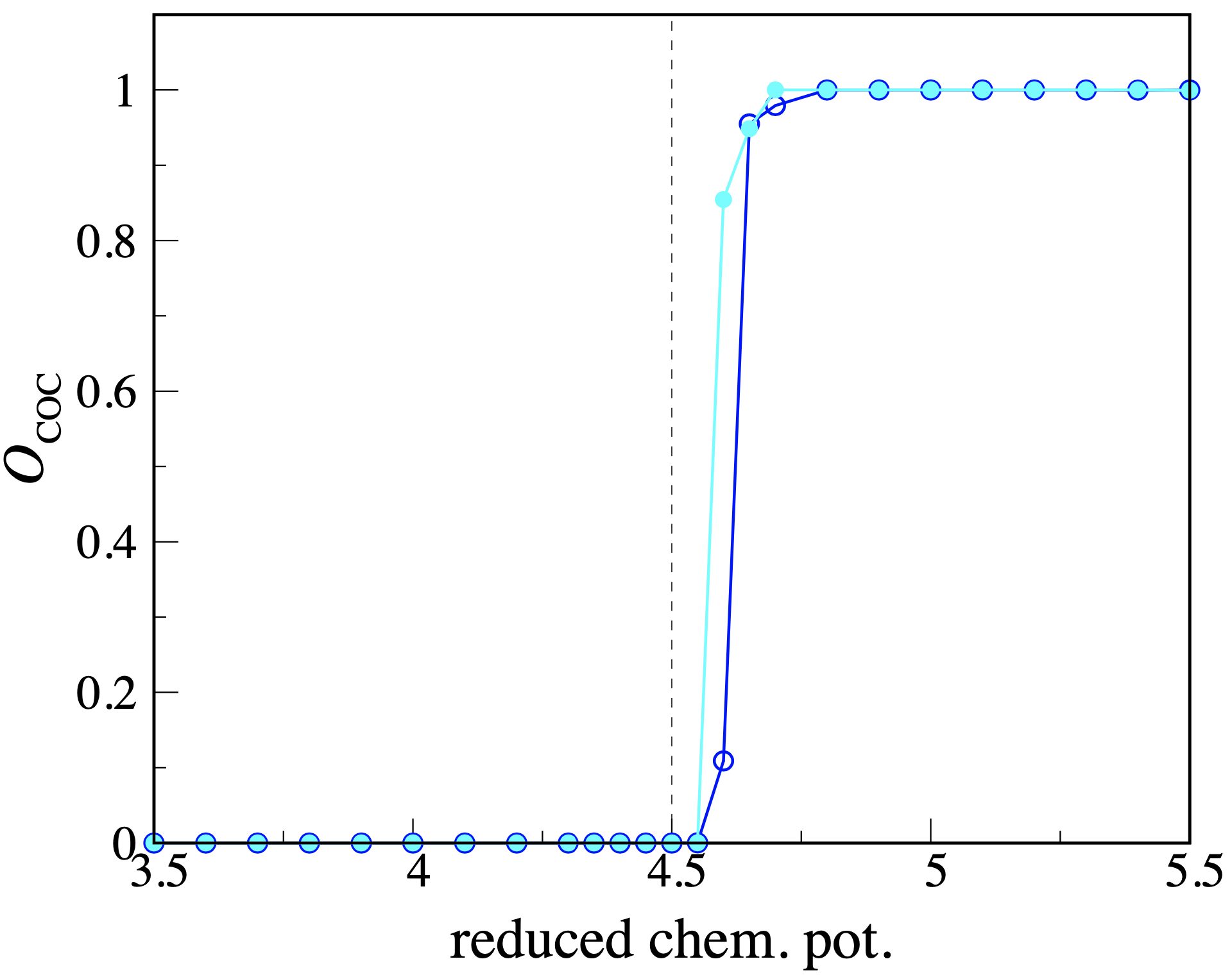}
\end{center}
\caption{Lattice gas on a PD mesh, for $\gamma=1$ and $T^*=0.05$: $N$ vs. $\mu$ across the transition between ICO+CUB and ICO+COC. We plot data from two sequences of runs, ascending (blue) and descending (cyan). The top left and top right panels differ for the number of equilibrium cycles performed in each run, which is $5\times 10^6$ and $5\times 10^7$, respectively. Bottom panels: OPs for ICO+CUB and ICO+COC, for the simulation with $5\times 10^7$ equilibrium cycles per run. We see a narrow interval of $\mu^*$ values around $9/2$ where the order is neither ICO+CUB nor ICO+COC.}
\label{fig5}
\end{figure}

%
%
\begin{figure}
\begin{center}
\includegraphics[width=6.5cm]{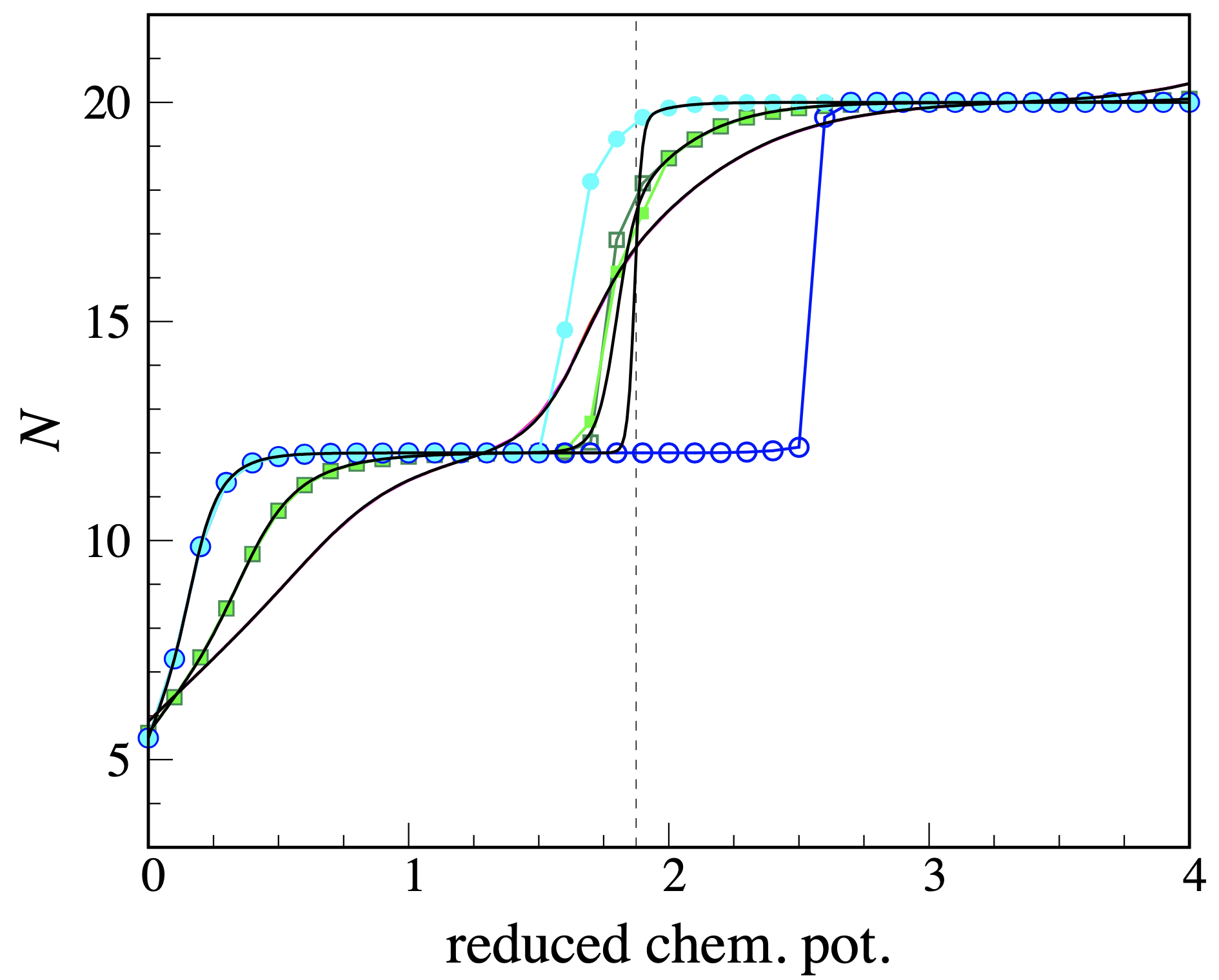}
\includegraphics[width=6.5cm]{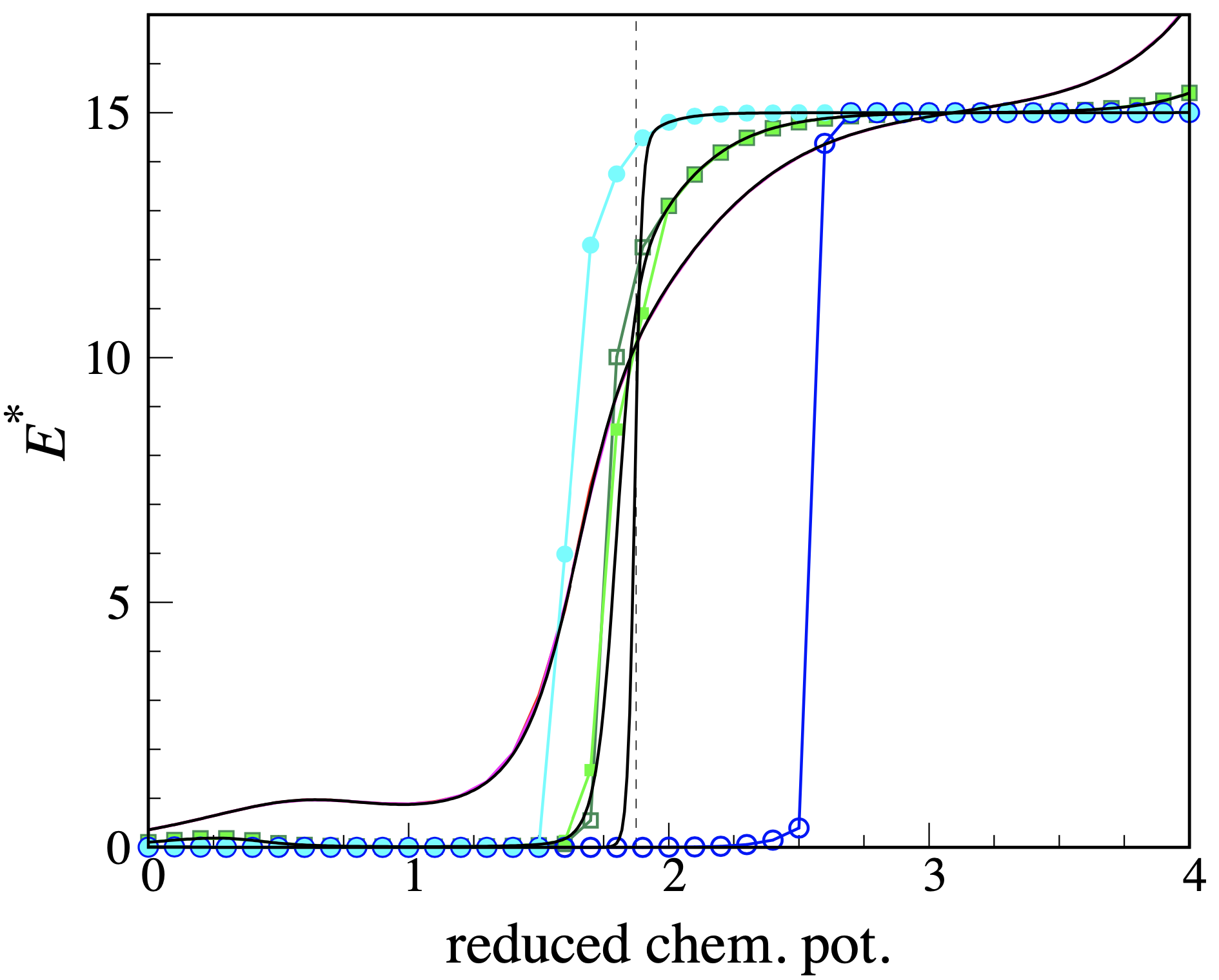}
\end{center}
\caption{MC results for $\gamma=1/2$ and $T^*=0.1,0.2$, and 0.3. The colored points and lines are the same MC data shown in Fig.\,3 ($T^*=0.1$, blue and cyan dots; $T^*=0.2$, emerald and green squares; $T^*=0.3$, red and pink). The superimposed black lines are the outcome of a Wang-Landau simulation. Left: average number of particles. Right: average energy.}
\label{fig6}
\end{figure}

%
%
\begin{figure}
\begin{center}
\includegraphics[width=6.5cm]{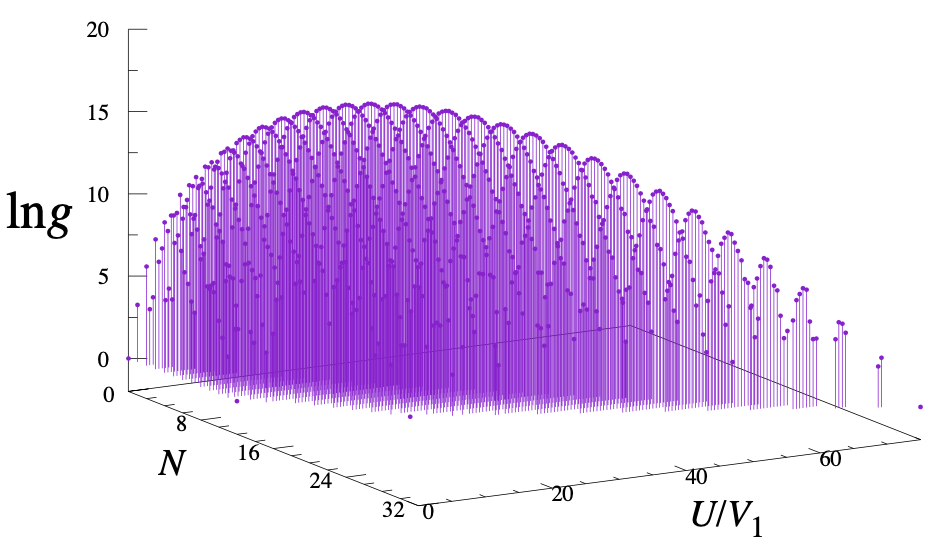}
\includegraphics[width=6.5cm]{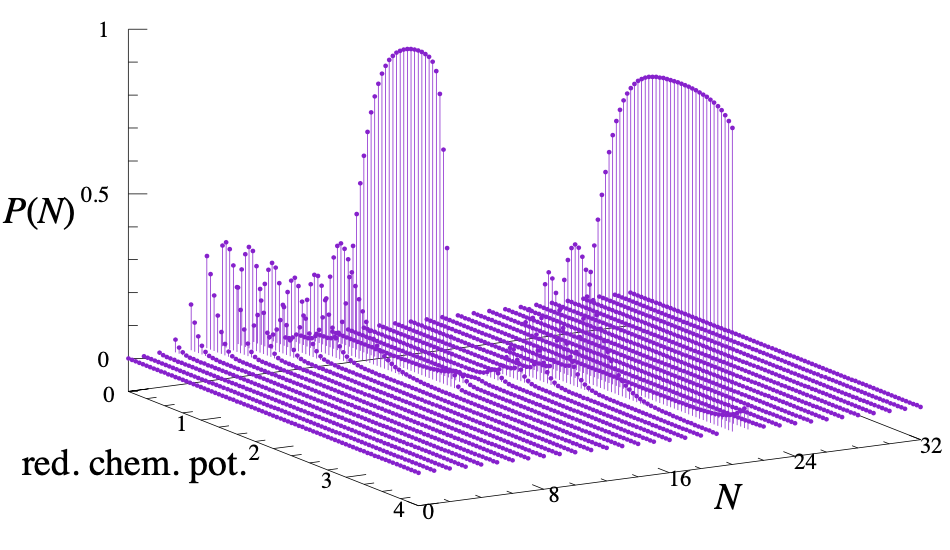}
\end{center}
\caption{Wang-Landau simulation for $\gamma=1/2$ and $T^*=0.2$. Left: logarithm of the density of states $g$, defined in terms of $N$ and $U$ (see text). The four points where $g=1$ correspond to the $T=0$ phases: ``empty'', perfect ICO, perfect DOD, and ``full''. Right: probability function of the particle number $N$, plotted as a function of the reduced chemical potential.}
\label{fig7}
\end{figure}

As anticipated, hysteresis is an annoying problem due to the inadequacy of Metropolis dynamics to overcome free-energy barriers. To solve this issue we have employed the Wang-Landau algorithm~\cite{Wang}, which directly computes the density of states and is thus particularly suited for a free-energy landscape with multiple minima. Compared to the original algorithm, the refinement parameter $\ln f$ was reduced at a slower rate (at regular intervals, $\ln f$ is divided by 1.1 rather than 2), which greatly reduces the (already small) saturation error~\cite{Belardinelli}. We report results for $\gamma=1/2$ in Figs.\,6 and 7. In the former figure, $N$ and $E^*$ are plotted as a function of $\mu$ across the ICO-DOD transition; we see that Metropolis sampling is indeed adequate for $T^*=0.3$ (notice, in particular, how the Wang-Landau data carefully interpolate the Metropolis data in the low-$\mu$ region, where a small bump is present in the energy). At lower temperatures, only the Wang-Landau simulation is unaffected by hysteresis. For completeness, for $T^*=0.2$ we plot in Fig.\,7 the density of states $g$ as a function of $N$ and $U$ (i.e., the sum of the first two terms on the r.h.s. of (\ref{eq-1})) and the probability density of the particle number, $P(N)$. In the ICO region, well before the transition at $\mu^*=15/8$, a second peak builds up in $P(N)$, which, as $\mu$ is increased, is gradually shifted to larger and larger $N$ values until becoming centered at $N=20$.

Finally, it is useful to compare MC results with the outcome of a mean-field (MF) theory. The simplest approach is to estimate the grand potential of (\ref{eq-1}) using the Gibbs-Bogoliubov (GB) inequality with a trial probability density $\pi[c]$ given as an uncorrelated product of one-site terms:
\be
\pi[c]=\prod_{i=1}^{32}\pi_i^{(1)}(c_i)
\label{eq-5}
\ee
with
\be
\pi_i^{(1)}(c)=\left\{
\begin{array}{ll}
\pi(c;\rho_{\rm A}) & \,,\,\,\,i\in{\rm A}\\
\pi(c;\rho_{\rm B}) & \,,\,\,\,i\in{\rm B}\\
\pi(c;\rho_{\rm C}) & \,,\,\,\,i\in{\rm C}\,.
\end{array}
\right.
\label{eq-6}
\ee
In the previous equation,
\be
\pi(c;\rho_{\rm A})=\rho_{\rm A}\delta_{c,1}+(1-\rho_{\rm A})\delta_{c,0}
\label{eq-7}
\ee
with $0\le\rho_{\rm A}\le 1$, and similarly for B and C. The rationale behind Eq.~(\ref{eq-6}) is that the average occupancy takes a possibly different value in each set of equivalent nodes, being $\rho_{\rm A}$ for the icosahedral set, $\rho_{\rm B}$ for the cubic set, and $\rho_{\rm C}$ for the co-cubic set.

An upper bound to the exact grand potential $\Omega$ is the GB grand potential $\Omega^*$,
\be
\Omega^*=\langle H\rangle+k_{\rm B}T\langle\ln\pi\rangle\,,
\label{eq-8}
\ee
where $\langle O[c]\rangle=\sum_{\{c\}}\pi[c]O[c]$. Then, it is a simple matter to show that
\ba
\Omega^*&=&24V_1\rho_{\rm A}\rho_{\rm B}+36V_1\rho_{\rm A}\rho_{\rm C}+24\gamma V_1\rho_{\rm B}\rho_{\rm C}+6\gamma V_1\rho_{\rm C}^2-12\mu\rho_{\rm A}-8\mu\rho_{\rm B}-12\mu\rho_{\rm C}
\nonumber \\
&&+12k_{\rm B}T\left[\rho_{\rm A}\ln\rho_{\rm A}+(1-\rho_{\rm A})\ln(1-\rho_{\rm A})\right]+8k_{\rm B}T\left[\rho_{\rm B}\ln\rho_{\rm B}+(1-\rho_{\rm B})\ln(1-\rho_{\rm B})\right]
\nonumber \\
&&+12k_{\rm B}T\left[\rho_{\rm C}\ln\rho_{\rm C}+(1-\rho_{\rm C})\ln(1-\rho_{\rm C})\right]\,.
\label{eq-9}
\ea
Observe that the value of $\Omega^*$ in the putative ground states listed in Sec.\,2.1 exactly reproduces their respective grand potentials. The stationary values of (\ref{eq-9}) fulfill the coupled equations
\ba
\rho_{\rm A}&=&\frac{1}{e^{\beta(2V_1\rho_{\rm B}+3V_1\rho_{\rm C}-\mu)}+1}\,,\,\,\,\rho_B=\frac{1}{e^{\beta(3V_1\rho_{\rm A}+3\gamma V_1\rho_{\rm C}-\mu)}+1}\,,\,\,\,{\rm and}
\nonumber \\
\rho_{\rm C}&=&\frac{1}{e^{\beta(3V_1\rho_{\rm A}+2\gamma V_1\rho_{\rm B}+\gamma V_1\rho_{\rm C}-\mu)}+1}\,.
\label{eq-10}
\ea
These equations are solved numerically, seeking the solution that provides the absolute minimum $\Omega^*$ for the given $T$ and $\mu$.

%
%
\begin{figure}
\begin{center}
\includegraphics[angle=-90,width=13cm]{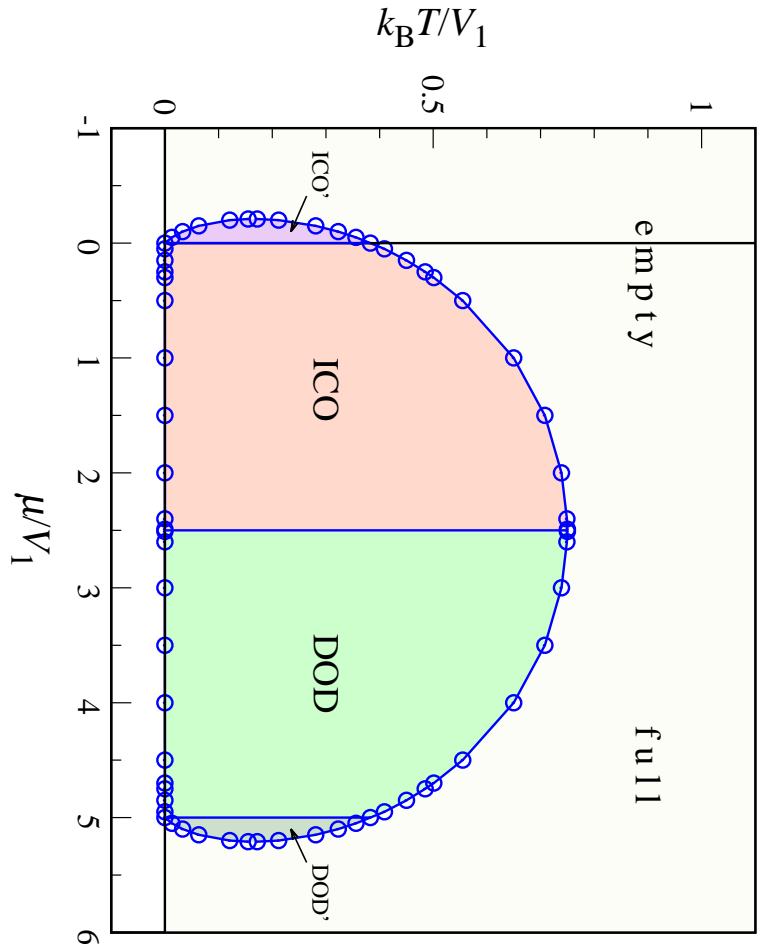}
\end{center}
\caption{Phase diagram of the lattice gas on a PD mesh as predicted by MF theory, for $\gamma=2/3$. All phase boundaries are first-order lines, see more in the text.}
\label{fig8}
\end{figure}

To have a flavor of how MF theory works, we draw in Fig.\,8 the theoretical phase diagram on the $\mu$-$T$ plane for $\gamma=2/3$, corresponding to a lattice gas where the icosahedral phase covers a $\mu$ range as wide as that of the dodecahedral phase, see Fig.\,2. Owing to a symmetry property of Eqs.\,(\ref{eq-10}), the phase diagram is symmetric around $\mu=5/2$ (see also Sec. 3.2). At $T=0$ the phase boundaries in Fig.\,8 are exact. As $T$ grows, the theory predicts a gradual weakening of ICO and DOD orders, as witnessed by the decrease of $|\rho_{\rm A}-\rho_{\rm B}|$ on heating, eventually resulting in an abrupt (first-order) transition to either ``empty'' ($\rho_{\rm A}=\rho_{\rm B}=\rho_{\rm C}<0.5$) or ``full'' ($\rho_{\rm A}=\rho_{\rm B}=\rho_{\rm C}>0.5$), according to whether $\mu<5/2$ or $\mu>5/2$. Clearly, this singularity is an artifact of MF theory, since no sharp transition is present in our system for $T>0$. Another unphysical prediction of the theory concerns the behavior of the model in a narrow strip of temperatures and chemical potentials near $\mu^*=0$ and $\mu^*=5$. In Fig.\,8 we have denoted ICO$^\prime$ a phase where $0.5>\rho_{\rm A}>\rho_{\rm B}=\rho_{\rm C}$ and DOD$^\prime$ a phase where $0.5<\rho_{\rm A}<\rho_{\rm B}=\rho_{\rm C}$. These two phases have no counterpart in the simulation, hence they are just an unwanted outcome of MF theory.

\section{Hard-core bosons on a spherical mesh: extended Bose-Hubbard model}

The PD mesh is regular enough that we can study the quantum analog of the lattice-gas model in relatively simple terms. The obvious bosonic counterpart of (\ref{eq-1}) is the hard-core limit of the extended Bose-Hubbard (BH) model
\ba
H&=&-t\sum_{\langle i,j\rangle}\left(a_i^\dagger a_j+a_j^\dagger a_i\right)-t\sum_{\langle\langle k,l\rangle\rangle}\left(a_k^\dagger a_l+a_l^\dagger a_k\right)
\nonumber \\
&&+\frac{U}{2}\sum_in_i(n_i-1)+V_1\sum_{\langle i,j\rangle}n_in_j+V_2\sum_{\langle\langle k,l\rangle\rangle}n_kn_l-\mu\sum_in_i\,,
\label{eq-11}
\ea
where $a_i,a_i^\dagger$ are bosonic field operators and $n_i=a_i^\dagger a_i$ is a number operator. Moreover, $t\ge 0$ is the hopping amplitude, taken for simplicity to be the same for first- and second-neighbor pairs, whereas $U>0$ is the on-site repulsion. In the hard-core limit $U\rightarrow +\infty$, the site occupancies are effectively restricted to zero or one and the $U$ term can thus be discarded. For hard-core bosons, creation and annihilation operators at different sites commute, whereas $a_i$ and $a_i^\dagger$ are {\em anticommuting} operators as a result of the dynamical suppression of Fock states with two or more particles in the same site~\cite{Morita}.

In the original BH model~\cite{Fisher,Rokhsar,Krauth}, where the second, fourth, and fifth terms in $H$ are absent, the tunneling term (kinetic energy) is minimized by a condensed state spread over the entire volume of the system, whereas the potential energy favors particle localization. As a result, the $T=0$ system exists as either a superfluid (large $t/U$) or a Mott insulating ground state (small $t/U$), separated by a quantum transition. The Bose-Hubbard Hamiltonian can be derived starting from the second-quantized Hamiltonian describing a gas of ultracold bosonic atoms subject to an optical-lattice potential~\cite{Jaksch}.

The phase scenario becomes richer when the range of interaction is increased: depending on the lattice, other Mott insulating ground states (density waves) may appear; moreover, crystalline order may coexist with superfluidity (supersolids)~\cite{Batrouni,vanOtterlo,Wessel,Kovrizhin,Pollet,Iskin}.
Supersolidity is a fascinating property of quantum matter, which has only recently been experimentally detected in a gas of dipolar atoms~\cite{Tanzi,Boettcher,Chomaz}. In a supersolid, atoms can simultaneously support frictionless flow and form a crystal. As suggested by Leggett, a rotating supersolid should have a moment of inertia that is reduced with respect to its classical value~\cite{Leggett}. This phenomenon is called ``nonclassical rotational inertia'' and its first observation is reported in a paper published this year~\cite{Tanzi2}.

Due to the semiregular character of the PD mesh, in our system the superfluid phase would be discouraged in favor of less-symmetric condensed phases, and a supersolid region will then occur at low temperature. To check this expectation, we employ a mean-field theory, an approach known to give accurate results in the continuum~\cite{Kunimi,Macri,Prestipino6}.

\subsection{Decoupling approximation}

As done in Refs.\,\cite{Prestipino5,Prestipino7}, we analyze the phase diagram of the extended BH model in the hard-core limit using the decoupling approximation (DA)~\cite{Sheshadri,Gheeraert}. The latter approach consists in linearizing the hopping and interaction terms in (\ref{eq-11}) as
\be
a_i^\dagger a_j\approx a_i^\dagger\left<a_j\right>+\big<a_i^\dagger\big>a_j-\big<a_i^\dagger\big>\left<a_j\right>\,\,\,\,\,\,{\rm and}\,\,\,\,\,\,n_in_j\approx n_i\left<n_j\right>+\left<n_i\right>n_j-\left<n_i\right>\left<n_j\right>\,,
\label{eq-12}
\ee
where the thermal averages $\left<a_i\right>\equiv\phi_i$ and $\left<n_i\right>\equiv\rho_i$ are determined self-consistently; $\phi_i$ and $\rho_i$ represent the superfluid OP and the average occupancy in the $i$-th site, respectively (the condensed fraction is $|\phi_i|^2$, see e.g. \cite{Prestipino7}). The DA Hamiltonian is a sum of one-site terms, given by
\ba
H_{\rm DA}&=&-t\sum_i\big(F_ia_i^\dagger+F_i^*a_i-F_i\phi_i^*\big)
\nonumber \\
&&+\frac{V_1}{2}\sum_i\left(2R_in_i-R_i\rho_i\right)+\frac{V_2}{2}\sum_i\left(2R'_in_i-R'_i\rho_i\right)-\mu\sum_in_i
\label{eq-13}
\ea
with $F_i=\sum_{j\in{\rm NN}_i,{\rm NNN}_i}\phi_j,R_i=\sum_{j\in{\rm NN}_i}\rho_j$, and $R'_i=\sum_{j\in{\rm NNN}_i}\rho_j$ (denoting NN$_i$ and NNN$_i$ the first and second neighbors of $i$, respectively). While referring to \cite{Prestipino7} for a full justification of DA, it is worth to underline that the self-consistency equations for $\phi_i$ and $\rho_i$ are also the conditions under which the grand potential of (\ref{eq-13}) is stationary. If more stationary solutions are found, we must select the one providing the minimum grand potential.

As discussed before, the 32 nodes of the PD mesh are naturally classified as icosahedral (A), cubic (B), or co-cubic (C), implying that the number of variational parameters in (\ref{eq-13}) is reduced to six. A phase with $\phi_{\rm A}=\phi_{\rm B}=\phi_{\rm C}=0$ is a Mott insulator, whereas a homogeneous occupancy together with $\phi_{\rm A}=\phi_{\rm B}=\phi_{\rm C}\ne 0$ defines a superfluid. Any unbalance between $\phi_{\rm A},\phi_{\rm B}$, and $\phi_{\rm C}$ corresponds to a supersolid.

Looking at Fig.\,1 we soon realize that
\ba
&&F_{\rm A}=2\phi_{\rm B}+3\phi_{\rm C}\,,\,\,\,F_{\rm B}=3\phi_{\rm A}+3\phi_{\rm C}\,,\,\,\,F_{\rm C}=3\phi_{\rm A}+2\phi_{\rm B}+\phi_{\rm C}\,;
\nonumber \\
&&R_{\rm A}=2\rho_{\rm B}+3\rho_{\rm C}\,,\,\,\,R_{\rm B}=3\rho_{\rm A}\,,\,\,\,R_{\rm C}=3\rho_{\rm A}\,;
\nonumber \\
&&R'_{\rm A}=0\,,\,\,\,R'_{\rm B}=3\rho_{\rm C}\,,\,\,\,R'_{\rm C}=2\rho_{\rm B}+\rho_{\rm C}\,,
\label{eq-14}
\ea
in such a way that the DA Hamiltonian becomes
\be
H_{\rm DA}=12h^{({\rm A})}+8h^{({\rm B})}+12h^{({\rm C})}
\label{eq-15}
\ee
with
\small
\ba
h^{({\rm A})}&=&E_0^{({\rm A})}-t\left[(2\phi_{\rm B}+3\phi_{\rm C})a_{\rm A}^\dagger+(2\phi_{\rm B}^*+3\phi_{\rm C}^*)a_{\rm A}\right]+(2V_1\rho_{\rm B}+3V_1\rho_{\rm C}-\mu)n_{\rm A}\,;
\nonumber \\
h^{({\rm B})}&=&E_0^{({\rm B})}-3t\left[(\phi_{\rm A}+\phi_{\rm C})a_{\rm B}^\dagger+(\phi_{\rm A}^*+\phi_{\rm C}^*)a_{\rm B}\right]+(3V_1\rho_{\rm A}+3V_2\rho_{\rm C}-\mu)n_{\rm B}\,;
\nonumber \\
h^{({\rm C})}&=&E_0^{({\rm C})}-t\left[(3\phi_{\rm A}+2\phi_{\rm B}+\phi_{\rm C})a_{\rm C}^\dagger+(3\phi_{\rm A}^*+2\phi_{\rm B}^*+\phi_{\rm C}^*)a_{\rm C}\right]+(3V_1\rho_{\rm A}+2V_2\rho_{\rm B}+V_2\rho_{\rm C}-\mu)n_{\rm C}
\nonumber \\
\label{eq-16}
\ea
\normalsize
and
\ba
E_0^{({\rm A})}&=&t\phi_{\rm A}^*(2\phi_{\rm B}+3\phi_{\rm C})-V_1\rho_{\rm A}\rho_{\rm B}-\frac{3}{2}V_1\rho_{\rm A}\rho_{\rm C}\,;
\nonumber \\
E_0^{({\rm B})}&=&3t\phi_{\rm B}^*(\phi_{\rm A}+\phi_{\rm C})-\frac{3}{2}V_1\rho_{\rm A}\rho_{\rm B}-\frac{3}{2}V_2\rho_{\rm B}\rho_{\rm C}\,;
\nonumber \\
E_0^{({\rm C})}&=&t\phi_{\rm C}^*(3\phi_{\rm A}+2\phi_{\rm B}+\phi_{\rm C})-\frac{3}{2}V_1\rho_{\rm A}\rho_{\rm C}-V_2\rho_{\rm B}\rho_{\rm C}-\frac{1}{2}V_2\rho_{\rm C}^2\,.
\label{eq-17}
\ea

In the hard-core limit, the eigenvalues of each partial Hamiltonian in (\ref{eq-16}) follow from the diagonalization of a $2\times 2$ matrix. We easily obtain:
\small
\ba
\lambda_\pm^{({\rm A})}&=&E_0^{({\rm A})}+\frac{2V_1\rho_{\rm B}+3V_1\rho_{\rm C}-\mu}{2}\pm\sqrt{\left(\frac{2V_1\rho_{\rm B}+3V_1\rho_{\rm C}-\mu}{2}\right)^2+t^2|2\phi_{\rm B}+3\phi_{\rm C}|^2}\,;
\nonumber \\
\lambda_\pm^{({\rm B})}&=&E_0^{({\rm B})}+\frac{3V_1\rho_{\rm A}+3V_2\rho_{\rm C}-\mu}{2}\pm\sqrt{\left(\frac{3V_1\rho_{\rm A}+3V_2\rho_{\rm C}-\mu}{2}\right)^2+9t^2|\phi_{\rm A}+\phi_{\rm C}|^2}\,;
\nonumber \\
\lambda_\pm^{({\rm C})}&=&E_0^{({\rm C})}+\frac{3V_1\rho_{\rm A}+2V_2\rho_{\rm B}+V_2\rho_{\rm C}-\mu}{2}
\nonumber \\
&&\pm\sqrt{\left(\frac{3V_1\rho_{\rm A}+2V_2\rho_{\rm B}+V_2\rho_{\rm C}-\mu}{2}\right)^2+t^2|3\phi_{\rm A}+2\phi_{\rm B}+\phi_{\rm C}|^2}\,.
\label{eq-18}
\ea
\normalsize
Therefore, for $T=0$ the grand potential of (\ref{eq-13}) reads
\ba
\Omega&=&12\lambda_-^{({\rm A})}+8\lambda_-^{({\rm B})}+12\lambda_-^{({\rm C})}
\nonumber \\
&=&E_0+6(2V_1\rho_{\rm B}+3V_1\rho_{\rm C}-\mu)+4(3V_1\rho_{\rm A}+3V_2\rho_{\rm C}-\mu)
\nonumber \\
&&+6(3V_1\rho_{\rm A}+2V_2\rho_{\rm B}+V_2\rho_{\rm C}-\mu)-6\sqrt{\circled{\rm A}}-4\sqrt{\circled{\rm B}}-6\sqrt{\circled{\rm C}}
\label{eq-19}
\ea
with $E_0=12E_0^{({\rm A})}+8E_0^{({\rm B})}+12E_0^{({\rm C})}$ and
\ba
\circled{\rm A}&=&(2V_1\rho_{\rm B}+3V_1\rho_{\rm C}-\mu)^2+4t^2|2\phi_{\rm B}+3\phi_{\rm C}|^2\,;
\nonumber \\
\circled{\rm B}&=&(3V_1\rho_{\rm A}+3V_2\rho_{\rm C}-\mu)^2+36t^2|\phi_{\rm A}+\phi_{\rm C}|^2\,;
\nonumber \\
\circled{\rm C}&=&(3V_1\rho_{\rm A}+2V_2\rho_{\rm B}+V_2\rho_{\rm C}-\mu)^2+4t^2|3\phi_{\rm A}+2\phi_{\rm B}+\phi_{\rm C}|^2\,.
\label{eq-20}
\ea
For $T>0$, the partition function of (\ref{eq-13}) reads:
\be
\Xi=\left(\sum_\pm e^{-\beta\lambda_\pm^{({\rm A})}}\right)^{12}\left(\sum_\pm e^{-\beta\lambda_\pm^{({\rm B})}}\right)^8\left(\sum_\pm e^{-\beta\lambda_\pm^{({\rm C})}}\right)^{12}\,,
\label{eq-21}
\ee
yielding the grand potential
\ba
\Omega&=&-k_{\rm B}T\ln\Xi=E_0+30V_1\rho_{\rm A}+12(V_1+V_2)\rho_{\rm B}+18(V_1+V_2)\rho_{\rm C}-16\mu
\nonumber \\
&&-12k_{\rm B}T\ln\left[2\cosh\left(\frac{1}{2}\beta\sqrt{\circled{\rm A}}\right)\right]-8k_{\rm B}T\ln\left[2\cosh\left(\frac{1}{2}\beta\sqrt{\circled{\rm B}}\right)\right]
\nonumber \\
&&-12k_{\rm B}T\ln\left[2\cosh\left(\frac{1}{2}\beta\sqrt{\circled{\rm C}}\right)\right]\,.
\label{eq-22}
\ea

In seeking the stationary solutions of (\ref{eq-22}), it can be assumed --- without loss of generality --- that $\phi_{\rm A},\phi_{\rm B},\phi_{\rm C}$ are real and positive~\cite{Prestipino5,Prestipino7}. Putting the derivative of (\ref{eq-22}) with respect to each free parameter equal to zero, and suitably rearranging the formulae, we arrive at the coupled equations:
\ba
\rho_{\rm A}&=&\frac{1}{2}-\tanh\left(\frac{1}{2}\beta\sqrt{\circled{\rm A}}\right)\frac{2V_1\rho_{\rm B}+3V_1\rho_{\rm C}-\mu}{2\sqrt{\circled{\rm A}}}\,;
\nonumber \\
\rho_{\rm B}&=&\frac{1}{2}-\tanh\left(\frac{1}{2}\beta\sqrt{\circled{\rm B}}\right)\frac{3V_1\rho_{\rm A}+3V_2\rho_{\rm C}-\mu}{2\sqrt{\circled{\rm B}}}\,;
\nonumber \\
\rho_{\rm C}&=&\frac{1}{2}-\tanh\left(\frac{1}{2}\beta\sqrt{\circled{\rm C}}\right)\frac{3V_1\rho_{\rm A}+2V_2\rho_{\rm B}+V_2\rho_{\rm C}-\mu}{2\sqrt{\circled{\rm C}}}\,;
\nonumber \\
\phi_{\rm A}&=&\tanh\left(\frac{1}{2}\beta\sqrt{\circled{\rm A}}\right)\frac{t(2\phi_{\rm B}+3\phi_{\rm C})}{\sqrt{\circled{\rm A}}}\,;
\nonumber \\
\phi_{\rm B}&=&\tanh\left(\frac{1}{2}\beta\sqrt{\circled{\rm B}}\right)\frac{3t(\phi_{\rm A}+\phi_{\rm C})}{\sqrt{\circled{\rm B}}}\,;
\nonumber \\
\phi_{\rm C}&=&\tanh\left(\frac{1}{2}\beta\sqrt{\circled{\rm C}}\right)\frac{t(3\phi_{\rm A}+2\phi_{\rm B}+\phi_{\rm C})}{\sqrt{\circled{\rm C}}}\,.
\label{eq-23}
\ea
The above equations can be solved numerically by, e.g, the method described in \cite{Prestipino5}. Upon combining the six equations (\ref{eq-23}) together, the following identities are easily derived:
\ba
&&\left(\rho_{\rm A}-\frac{1}{2}\right)^2+\phi_{\rm A}^2=\frac{1}{4}\tanh^2\left(\frac{1}{2}\beta\sqrt{\circled{\rm A}}\right)\,;
\nonumber \\
&&\left(\rho_{\rm B}-\frac{1}{2}\right)^2+\phi_{\rm B}^2=\frac{1}{4}\tanh^2\left(\frac{1}{2}\beta\sqrt{\circled{\rm B}}\right)\,;
\nonumber \\
&&\left(\rho_{\rm C}-\frac{1}{2}\right)^2+\phi_{\rm C}^2=\frac{1}{4}\tanh^2\left(\frac{1}{2}\beta\sqrt{\circled{\rm C}}\right)\,,
\label{eq-24}
\ea
indicating that the value of each superfluid OP is comprised between 0 and $1/2$.

Before moving to numerical results, we show that for $t=0$ the DA theory is equivalent to the MF theory described in Section 2.2. Indeed, for $t=0$ the grand potential (\ref{eq-22}) becomes
\ba
\Omega&=&E_0-12k_{\rm B}T\ln\left[1+e^{-\beta(2V_1\rho_{\rm B}+3V_1\rho_{\rm C}-\mu)}\right]-8k_{\rm B}T\ln\left[1+e^{-\beta(3V_1\rho_{\rm A}+3V_2\rho_{\rm C}-\mu)}\right]
\nonumber \\
&&-12k_{\rm B}T\ln\left[1+e^{-\beta(3V_1\rho_{\rm A}+2V_2\rho_{\rm B}+V_2\rho_{\rm C}-\mu)}\right]
\label{eq-25}
\ea
with
\be
E_0=-24V_1\rho_{\rm A}\rho_{\rm B}-36V_1\rho_{\rm A}\rho_{\rm C}-24V_2\rho_{\rm B}\rho_{\rm C}-6V_2\rho_{\rm C}^2\,.
\label{eq-26}
\ee
Upon differentiating (\ref{eq-25}) with respect to each density parameter and putting the result equal to zero, the same equations (\ref{eq-10}) are eventually obtained. If these equations are substituted back into (\ref{eq-25}), then the grand potential (\ref{eq-9}) is obtained, indicating that the DA phase diagram of the $t=0$ quantum system is exactly identical to the phase diagram of the lattice-gas system in the MF approximation.

\subsection{Numerical results}

Until now, the value of $V_2$ was arbitrary. For the sake of example, we henceforth take $\gamma=V_2/V_1=2/3$. We note that the case $\gamma=1$ was considered in \cite{Prestipino7}.

%
%
\begin{figure}
\begin{center}
\includegraphics[angle=-90,width=13cm]{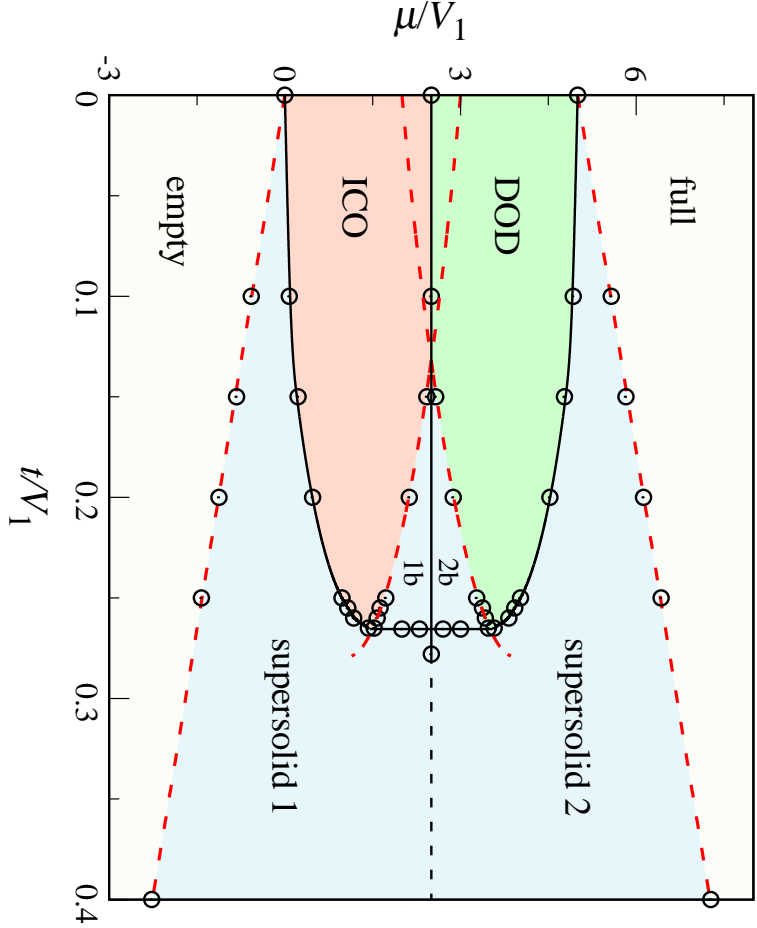}
\end{center}
\caption{DA phase diagram of the extended BH model at $T=0$, for $\gamma=2/3$. The open dots mark transition points. The dashed red curves are the continuous-transition loci derived in the text (cf. Eqs.~(\ref{eq-27})). The black dashed line marks the passing from supersolid 1 to 2 for $t\gtrsim 0.278$. Here the system is superfluid (see right panel of Fig.\,10). The remaining black lines represent first-order transitions.}
\label{fig9}
\end{figure}

We have first solved Eqs.~(\ref{eq-23}) numerically for $T=0$ and various $\mu$ values, being careful that the minimum $\Omega$ solution is picked out in each case. The resulting phase diagram is shown in Fig.\,9. In addition to the ``empty'' phase ($\rho_{\rm A}=\rho_{\rm B}=\rho_{\rm C}=0$) and the ``full'' phase ($\rho_{\rm A}=\rho_{\rm B}=\rho_{\rm C}=1$), we observe an icosahedral phase ($\rho_{\rm A}=1,\rho_{\rm B}=\rho_{\rm C}=0$) and a dodecahedral phase ($\rho_{\rm A}=0,\rho_{\rm B}=\rho_{\rm C}=1$). All these phases are insulating ($\phi_{\rm A}=\phi_{\rm B}=\phi_{\rm C}=0$) and incompressible (i.e., the density is constant throughout the phase). Upon increasing $t$ at fixed $\mu$ a supersolid phase invariably appears, characterized by $\rho_{\rm A}\ne\rho_{\rm B}=\rho_{\rm C}$. We also note that our phase diagram is symmetric around $\mu=5/2$. Indeed, we see from Eqs.\,(\ref{eq-23}) that $\rho_{\rm A,B}(\mu)=1-\rho_{\rm A,B}(5-\mu)$ and $\phi_{\rm A,B}(\mu)=\phi_{\rm A,B}(5-\mu)$, and an analogous symmetry property holds for (\ref{eq-22}). It is worth stressing the similarities and differences between Fig.\,9 and the phase diagram of hard-core bosons on a triangular lattice~\cite{Wessel,Zhang,Gheeraert}: The overall structure is the same, but the nature of the condensed phase at large $t$ is different, being herein supersolid rather than superfluid --- owing to the frustration effect associated with the existence of inequivalent nodes.

%
%
\begin{figure}
\begin{center}
\includegraphics[width=6.6cm]{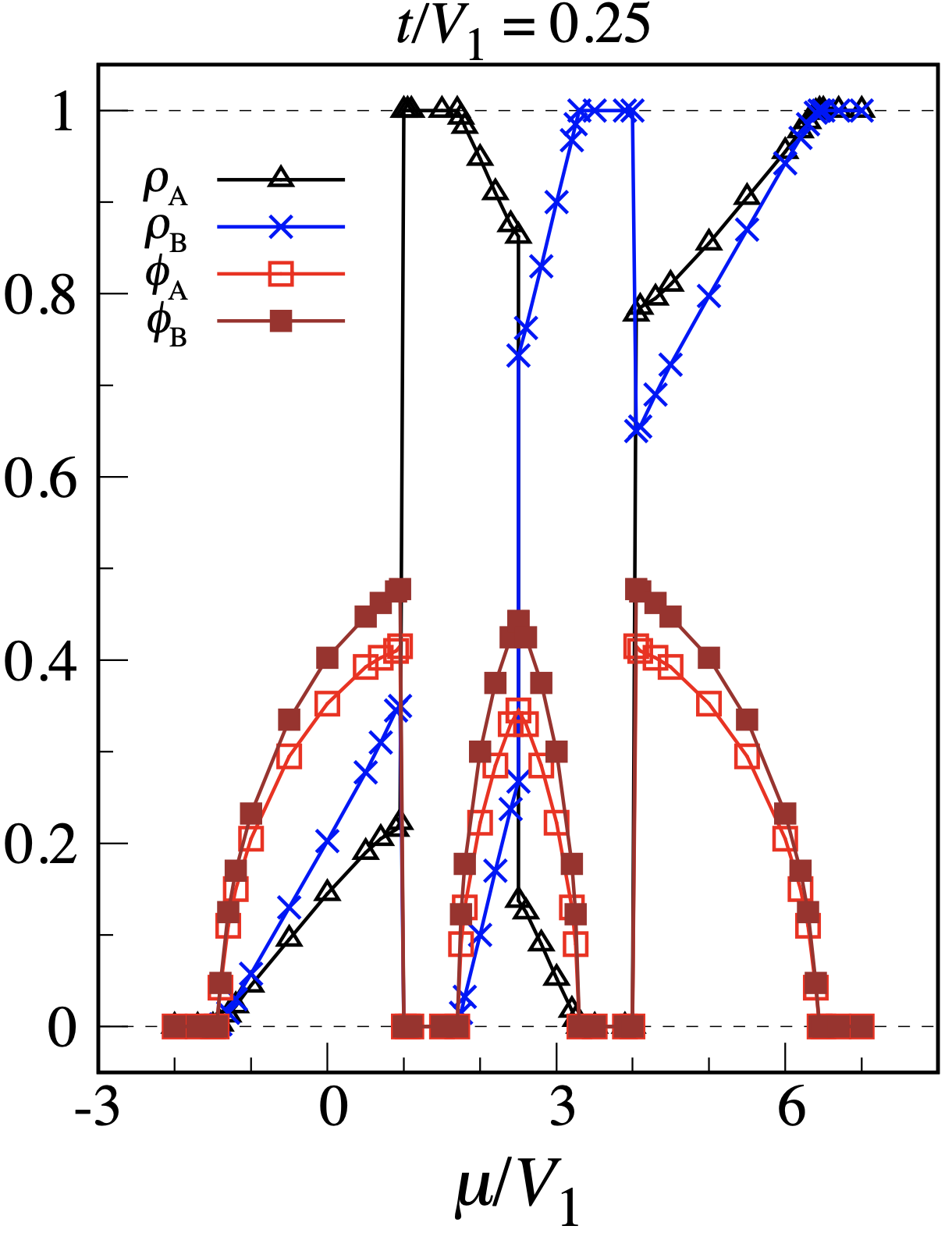}
\includegraphics[width=6.6cm]{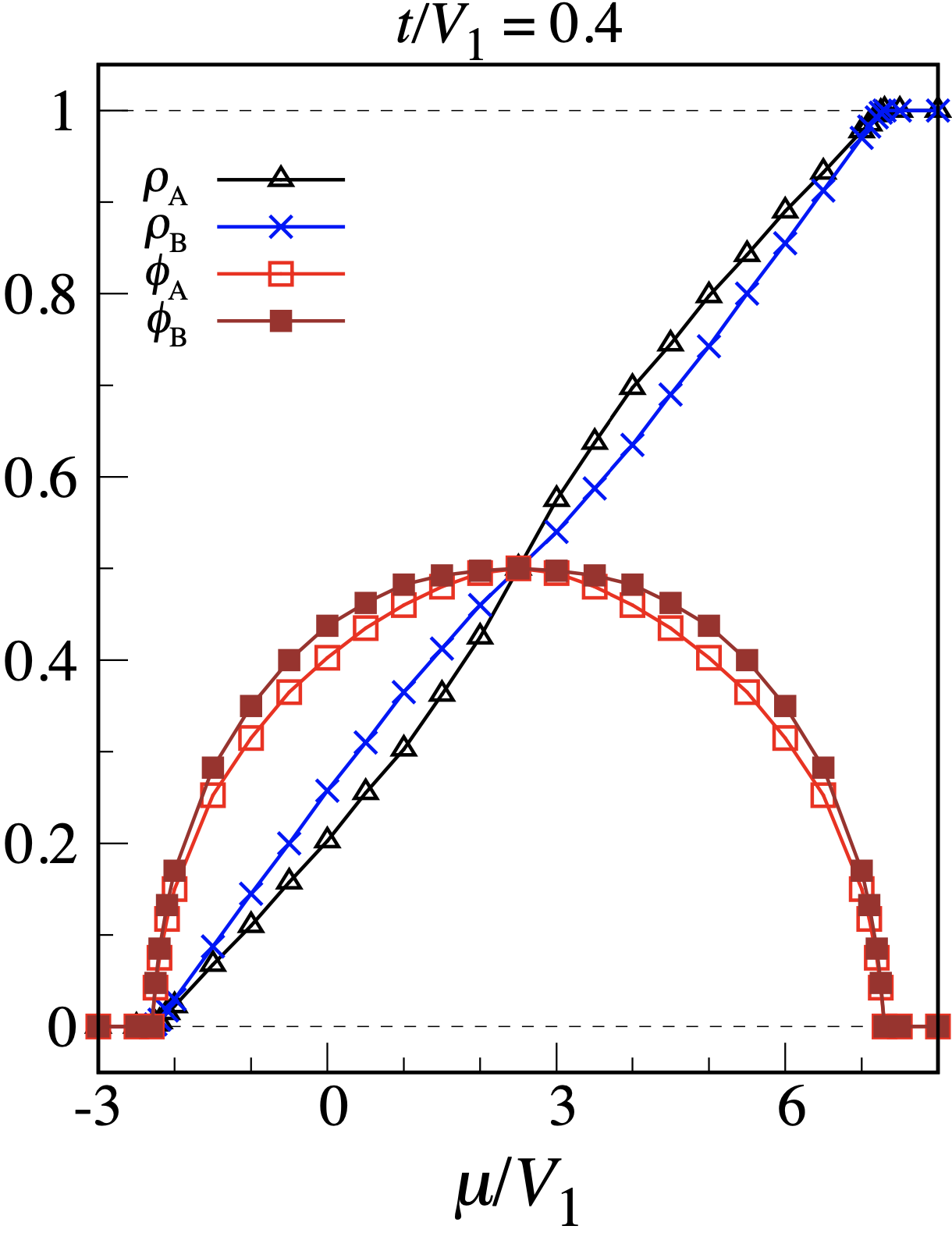}
\end{center}
\caption{DA results for $\gamma=2/3$. The OPs are plotted as a function of $\mu$ for fixed $t$ (left, $t/V_1=0.25$; right, $t/V_1=0.4$).}
\label{fig10}
\end{figure}

We actually distinguish four different supersolid phases (see Fig.\,10, where the OPs are plotted for two representative values of $t$). While $\phi_{\rm B}$ is always slightly larger than $\phi_{\rm A}$, in the region between the ICO and DOD lobes (where, in particular, $t<0.266$) we find $\rho_{\rm A}>\rho_{\rm B}$ for $\mu<5/2$ (supersolid 1b) and $\rho_{\rm A}<\rho_{\rm B}$ for $\mu>5/2$ (supersolid 2b). Outside the lobes, we instead find $\rho_{\rm A}<\rho_{\rm B}$ for $\mu<5/2$ (supersolid 1) and $\rho_{\rm A}>\rho_{\rm B}$ for $\mu>5/2$ (supersolid 2), namely the densities are in reverse order with respect to the reference ``solid'' phase. For $t\lesssim 0.278$, the values of $\rho_{\rm A}$ and $\rho_{\rm B}$ jump discontinuously at $\mu=5/2$, signaling that the phase transitions along this line are first-order. A further first-order line runs vertically near $t=0.266$, separating the supersolid phases 1b and 2b from the supersolid phases 1 and 2, respectively. Finally, there are four second-order transition lines: the two lines separating ``empty" and ``full'' from the adjacent supersolid, the descending part of the boundary between ICO and supersolid 1b, and the ascending part of the boundary between supersolid 2b and DOD.

Imposing B-C symmetry, we may simplify Eqs.\,(\ref{eq-23}) and then determine the equations for the continuous-transition loci, following the same procedure as illustrated in Ref.\,\cite{Prestipino7}. We eventually find:
\ba
\mu&=&-\frac{3+\sqrt{69}}{2}t\qquad\qquad\qquad\qquad\qquad({\rm ``empty"\text{-}supersolid\,\,boundary})\,;
\nonumber \\
\mu&=&5V_1+\frac{3+\sqrt{69}}{2}t\qquad\qquad\qquad\qquad({\rm ``full"\text{-}supersolid\,\,boundary})\,;
\nonumber \\
\mu&=&\frac{3V_1-3t\pm\sqrt{9V_1^2-18V_1t-51t^2}}{2}\,\,\,({\rm ICO\text{-}supersolid\,\,boundary})\,;
\nonumber \\
\mu&=&\frac{7V_1+3t\pm\sqrt{9V_1^2-18V_1t-51t^2}}{2}\,\,\,({\rm DOD\text{-}supersolid\,\,boundary})\,.
\label{eq-27}
\ea
In fact, looking at Fig.\,9 we see that the lower branch of the ICO-supersolid locus is only virtual, since this transition is preempted by a first-order phase transition. A similar comment applies for the upper branch of the DOD-supersolid locus.

%
%
\begin{figure}
\begin{center}
\includegraphics[angle=-90,width=13cm]{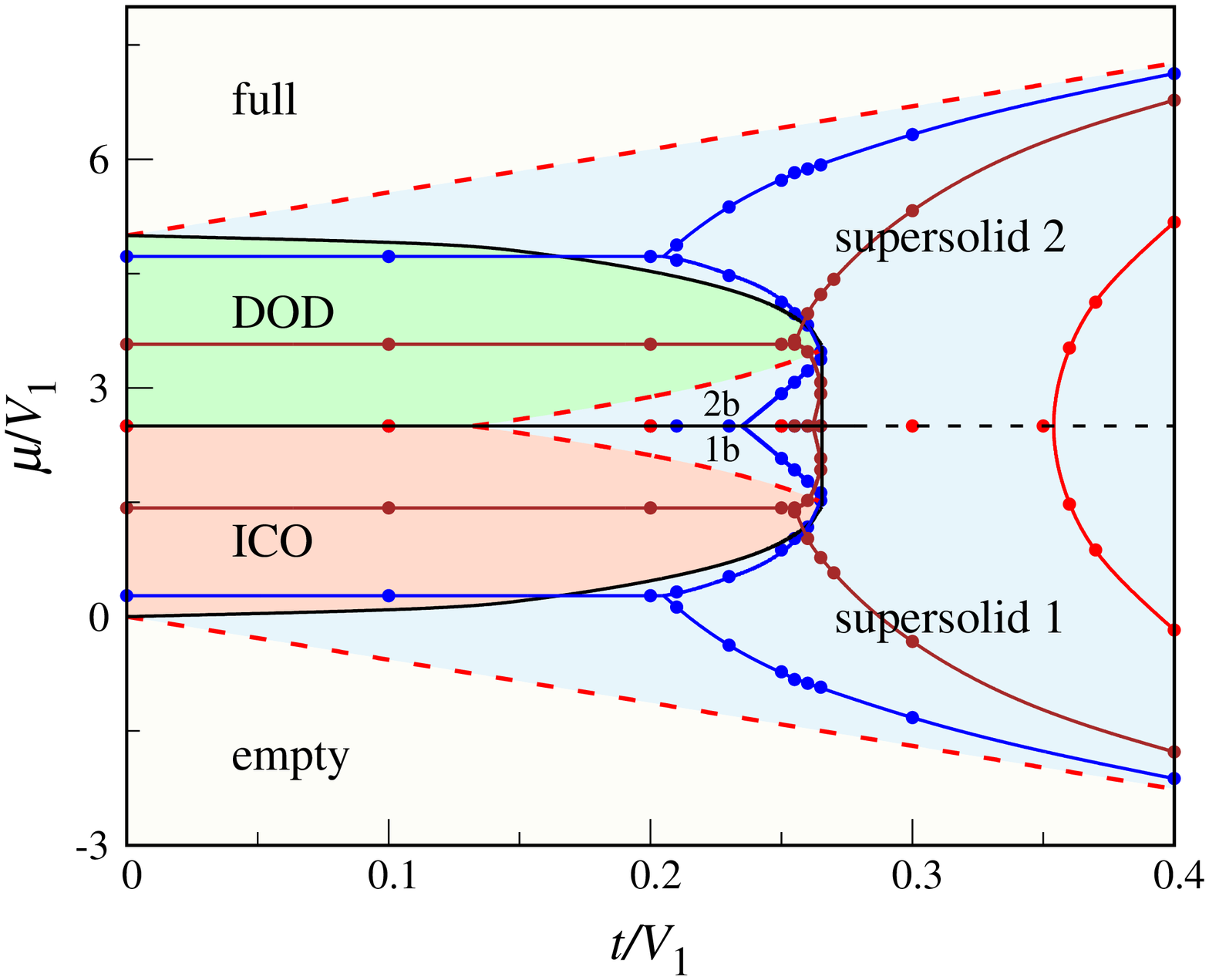}
\end{center}
\caption{DA phase diagram of the extended BH model for $\gamma=2/3$, plotted for three reduced temperatures $T^*$ (0.5, blue dots; 0.7, brown dots; 1, red dots). Lines through the points are drawn as a guide to the eye. For the sake of comparison, in the figure we have also reported the transition lines and phases for $T=0$.}
\label{fig11}
\end{figure}

For $T>0$ the phase diagram evolves in the way illustrated in Fig.\,11. Clearly, the indications of DA for non-zero temperatures are less accurate; moreover, the prediction of sharp phase boundaries is an artifact of the approximation, the transitions being actually smooth crossovers. Already for $T^*=0.5$ we observe a retreat of every supersolid phase with respect to $T=0$. The smallest $t$ value for which the system can be supersolid is now slightly larger than $0.20$. ICO and DOD too lose ground in favor of ``empty'' and ``full'', respectively, a trend that will become more marked on increasing $T$ further. An effect of finite temperature is that the occupation unbalance between A and B/C is no longer sharp in ICO and DOD, and is moreover $\mu$-dependent; but, similarly to $T=0$, the site occupancies are independent of $t$. Notice that the ICO and DOD densities as well as the ranges of stability are exactly the same as predicted by the MF theory of Sec.\,2.2 (see Fig.\,8). For any $T>0$ the densities are $\mu$-dependent and independent of $t$ also in the former ``empty'' and ``full'' phases, though still homogeneous throughout the mesh. For $T^*=0.7$ the supersolid 1b and 2b phases are nearly disappeared; furthermore, the $\mu$ extent of ICO and DOD is more than halved with respect to $T=0$. Finally, in the phase diagram for $T^*=1$ not only 1b and 2b but also ICO and DOD are no longer present. Moreover, the supersolid sector lies to the right of $t=0.35$; for $t<0.35$, the occupancies evolve continuously through $\mu^*=5/2$.

In spite of the elementary character of the DA, the main traits of the thermal evolution outlined above are correct, being in line with other theoretical studies~\cite{vanOosten,Buonsante,Lu} and simulations~\cite{Capogrosso-Sansone,Mahmud,Fang}. To recapitulate, the quantum phases are weakened by the thermal fluctuations associated with finite temperatures. Thus, for $T>0$ a normal fluid appears in the system. Here, the superfluid OP is zero and the density at each site of the mesh becomes non-integer. This is to be contrasted with the incompressibility of insulating quantum phases, which have integer occupancy at each site. Thermal fluctuations also undermine the supersolid phases, which are shifted towards higher and higher hopping amplitudes as $T$ is progressively increased.

\section{Conclusions}

Using a combination of analytic calculations and numerical experiments we have worked out the phase behavior of a lattice-gas model on the skeleton of the PD. Only particles connected by a PD edge are allowed to mutually interact, with different couplings for short and long edges. Depending on the ratio $\gamma$ between these couplings, various ordered phases are observed at $T=0$ (in addition to ``empty'' and ``full''): icosahedral, dodecahedral, icosahedral+cubic, and icosahedral+co-cubic. For $T>0$ we study the phase behavior of the lattice gas by MC simulation. The total number of sites (32) is sufficiently small that the system is quickly equilibrated at any temperature, with negligible uncertainties on the thermal properties. Despite the absence of sharp phase transitions in a finite system, at low temperature we observe strong hysteresis at some of these boundaries. We have shown that hysteresis, which would occur with any MC algorithm with local updates, can be cured by making an entropic sampling. In this respect, it would be intriguing to examine whether anything similar to the concept of nucleation barrier~\cite{Prestipino8,Prestipino9} applies for this model (but we leave this for future work).

A variation on the theme of the present model is one where the occupancy of sites is unrestricted. In this case, for high chemical potentials we expect to observe the formation of cluster phases, as in \cite{Franzini}. A mean-field theory similar to that formulated in \cite{Prestipino10} would probably suffice to obtain accurate predictions for the phase behavior of this system, thus making the simulation unnecessary.

Finally, we have considered the quantum analog of the lattice-gas model and solved it using the decoupling approximation. This theory predicts various Mott insulating ground states, each being the counterpart of a phase of the classical model, as well as a number of supersolids for higher hopping amplitudes. Admittedly, it is the frustration effect due to the semiregular character of the PD mesh that causes the superfluid phase to be superseded by a supersolid phase. The take-home message is that confining bosonic particles in a semiregular mesh is an easy way to stabilize a supersolid in an ultracold quantum gas. Upon heating, the extent of all the $T=0$ phases gets progressively reduced, leaving room to normal-fluid behavior.

\acknowledgments{We express our gratitude to one of the Referee for suggesting entropic sampling as a cure to the hysteresis apparent in Figs.\,3 and 4.}


\appendix
\section{Calculation of specific heats}
\setcounter{equation}{0}
\renewcommand{\theequation}{A\arabic{equation}}

In this Appendix, we derive the statistical-mechanical formulae for the specific heats in Eq.\,(\ref{eq-3}).

Working in the grand-canonical ensemble, the partition function (a sum over microstates) reads
\be
\Xi=\sum_\sigma e^{\beta\mu N_\sigma}e^{-\beta E_\sigma}\,,
\label{a-1}
\ee
with $T,V$, and $\mu$ as control parameters ($V$ is the system volume). The grand potential $\Omega$, that is the thermodynamic potential in the $T,V,\mu$ representation, is given by $\Omega=-k_{\rm B}T\ln\Xi$. The constant-$\mu$ specific heat is
\be
C_\mu=\frac{T}{N}\left.\frac{\partial S}{\partial T}\right|_{V,\mu}=-\frac{T}{N}\left.\frac{\partial^2\Omega}{\partial T^2}\right|_{V,\mu}\,,
\label{a-2}
\ee
where the number of particles $N$ is
\be
N=-\left.\frac{\partial\Omega}{\partial\mu}\right|_{T,V}=\left.\frac{\partial\ln\Xi}{\partial\beta\mu}\right|_{\beta,V}=\frac{\sum_\sigma N_\sigma e^{\beta\mu N_\sigma}e^{-\beta E_\sigma}}{\sum_\sigma e^{\beta\mu N_\sigma}e^{-\beta E_\sigma}}\equiv\langle N\rangle\,.
\label{a-3}
\ee
From
\be
\frac{\partial}{\partial T}=-k_{\rm B}\beta^2\frac{\partial}{\partial\beta}
\label{a-4}
\ee
it soon follows that
\be
C_\mu=\frac{1}{\langle N\rangle}k_{\rm B}\beta^2\left.\frac{\partial^2\ln\Xi}{\partial\beta^2}\right|_{V,\mu}\,.
\label{a-5}
\ee
Upon inserting Eq.\,(\ref{a-1}) into (\ref{a-5}) we eventually obtain:
\ba
\frac{C_\mu}{k_{\rm B}}&=&\frac{\beta^2}{\langle N\rangle}\left(\langle(E-\langle E\rangle)^2\rangle-2\mu(\langle EN\rangle-\langle E\rangle\langle N\rangle)+\mu^2\langle(N-\langle N\rangle)^2\rangle\right)
\nonumber \\
&\equiv&\frac{\beta^2}{\langle N\rangle}\left(\langle\delta E^2\rangle-2\mu\langle\delta E\delta N\rangle+\mu^2\langle\delta N^2\rangle\right)\,.
\label{a-6}
\ea
The averages in (\ref{a-6}) can be computed in a grand-canonical MC simulation.

Next, we focus on a different specific heat, calculated by keeping $N$ (and $V$) fixed:
\be
C_N=\frac{T}{N}\left.\frac{\partial S}{\partial T}\right|_{V,N}\,.
\label{a-7}
\ee
To enforce $N={\rm const.}$, the chemical potential must be in a suitable relation with $T$ and $V$:
\be
N(T,V,\mu)={\rm const.}\,\,\,\Longrightarrow\,\,\,\mu=\mu(T,V)\,.
\label{a-8}
\ee
Hence
\be
\left.\frac{\partial\mu}{\partial T}\right|_V=-\frac{\left(\frac{\partial N}{\partial T}\right)_{V,\mu}}{\left(\frac{\partial N}{\partial\mu}\right)_{T,V}}
\label{a-9}
\ee
and
\be
\left.\frac{\partial S}{\partial T}\right|_{V,N}=\left.\frac{\partial}{\partial T}S(T,V,\mu(T,V))\right|_V=\left.\frac{\partial S}{\partial T}\right|_{V,\mu}+\left.\frac{\partial S}{\partial\mu}\right|_{T,V}\left.\frac{\partial\mu}{\partial T}\right|_V\,.
\label{a-10}
\ee
In turn, the $\mu$ derivative of $S$ is given by a Maxwell relation:
\be
\left.\frac{\partial S}{\partial\mu}\right|_{T,V}=\left.\frac{\partial N}{\partial T}\right|_{V,\mu}=-\left.\frac{\partial N}{\partial\mu}\right|_{T,V}\left.\frac{\partial\mu}{\partial T}\right|_V\,.
\label{a-11}
\ee
Putting Eqs.\,(\ref{a-7})-(\ref{a-11}) together, we end up with:
\be
C_N=C_\mu-\frac{T}{\langle N\rangle}\frac{\left(\frac{\partial N}{\partial T}\right)_{V,\mu}^2}{\left(\frac{\partial N}{\partial\mu}\right)_{T,V}}\,.
\label{a-12}
\ee
Now, using (\ref{a-3}) we obtain:
\be
\left(\frac{\partial N}{\partial\mu}\right)_{T,V}=\beta\langle\delta N^2\rangle
\label{a-13}
\ee
and
\be
\left(\frac{\partial N}{\partial T}\right)_{V,\mu}=k_{\rm B}\beta^2\left(\langle\delta E\delta N\rangle-\mu\langle\delta N^2\rangle\right)\,.
\label{a-14}
\ee
Finally, substituting Eqs.\,(\ref{a-6}), (\ref{a-13}), and (\ref{a-14}) into (\ref{a-12}), we arrive at
\be
\frac{C_N}{k_{\rm B}}=\frac{\beta^2}{\langle N\rangle}\left(\langle\delta E^2\rangle-\frac{\langle\delta E\delta N\rangle^2}{\langle\delta N^2\rangle}\right)\,.
\label{a-15}
\ee

\section{Order parameters}
\setcounter{equation}{0}
\renewcommand{\theequation}{B\arabic{equation}}

We hereby introduce a few quantities, to be computed within the simulation of the lattice gas on a PD mesh, allowing us to identify the order present in the system at fixed $T$ and $\mu$. Clearly, there is no unique way to define these OPs --- our proposal below is just one possibility.

For our definition we need the current number of particles, ${\cal N}=\sum_ic_i$, and of occupied dodecahedral sites, ${\cal N}_{\rm D}=\sum_{i\in{\rm B}\cup{\rm C}}c_i$. A quantity sensitive to dodecahedral order is then obtained as follows. Let ${\cal O}_{\rm D}$ be ${\cal N}_{\rm D}/20$ if $19\le{\cal N}\le 21$ and ${\cal N}_{\rm D}\ge 19$, and 0 otherwise. Then, $O_{\rm DOD}=\langle{\cal O}_{\rm D}\rangle$. Similarly, to measure the amount of icosahedral order we compute a quantity ${\cal O}_{\rm I}$, defined as $({\cal N}-{\cal N}_{\rm D})/12$ if $11\le{\cal N}\le 13$ and ${\cal N}_{\rm D}\le 1$, and 0 otherwise. Then, $O_{\rm ICO}=\langle{\cal O}_{\rm I}\rangle$.

Measuring the degree of ICO+CUB order is more subtle, since in the phase region where this ``phase'' is stable the simulated system circulates, even at the lowest temperatures, between five different basins of microstates. As a result, in a long simulation the average occupancy will be the same for all dodecahedral sites. Here, the crucial observation is that the cosine of the angle $\theta_{ij}$ formed by the vector radii relative to two distinct cubic sites $i$ and $j$ is either $-1$ or $\pm 1/3$. In view of this, let ${\cal O}_{\rm CUB}$ be defined as $1-(1/3)\sum_{i,j\in{\rm B}\cup{\rm C}}^\prime c_ic_j(\cos\theta_{ij}+1)(\cos^2\theta_{ij}-1/9)^2$ if $19\le{\cal N}\le 21$ and $7\le{\cal N}_{\rm D}\le 9$, and 0 otherwise (the sum is over all distinct pairs of dodecahedral sites; the sum prefactor is just a reasonable choice). Then, $O_{\rm ICO+CUB}=\langle{\cal O}_{\rm CUB}\rangle$. The amount of ICO+COC order is similarly defined. Let ${\cal O}_{\rm COC}$ be $1-(1/3)\sum_{i,j\in{\rm B}\cup{\rm C}}^\prime(1-c_i)(1-c_j)(\cos\theta_{ij}+1)(\cos^2\theta_{ij}-1/9)^2$ if $23\le{\cal N}\le 25$ and $11\le{\cal N}_{\rm D}\le 13$, and 0 otherwise. Then, $O_{\rm ICO+COC}=\langle{\cal O}_{\rm COC}\rangle$.

\end{document}